\documentclass{ieeeoj} 
\usepackage{fix-cm}
\usepackage{amsfonts} \usepackage[caption=false,font=normalsize,labelfont=sf,textfont=sf,position=top]{subfig} \usepackage{textcomp} \usepackage{stfloats} \usepackage{cite} \usepackage{xcolor,soul,framed} \usepackage{amsmath,amssymb,amsfonts} \usepackage{array} \usepackage{mdwmath} \usepackage{mdwtab} \usepackage{eqparbox} \usepackage{url} \usepackage{tabularx} \usepackage{orcidlink} 
\usepackage{tikz} \usepackage{booktabs} \usepackage{graphicx}\usetikzlibrary{arrows.meta} \usepackage{multirow} \usepackage{siunitx} \usepackage{makecell} \usepackage{algorithm} \usepackage[noend]{algpseudocode} \usepackage{mathtools} \usepackage{placeins} \usepackage{nccmath} \usepackage{relsize}
\def\BibTeX{{\rm B\kern-.05em{\sc i\kern-.025em b}\kern-.08em
    T\kern-.1667em\lower.7ex\hbox{E}\kern-.125emX}}
\AtBeginDocument{\definecolor{ojcolor}{cmyk}{0.93,0.59,0.15,0.02}}
\def\OJlogo{\vspace{-14pt}\includegraphics[height=28pt]{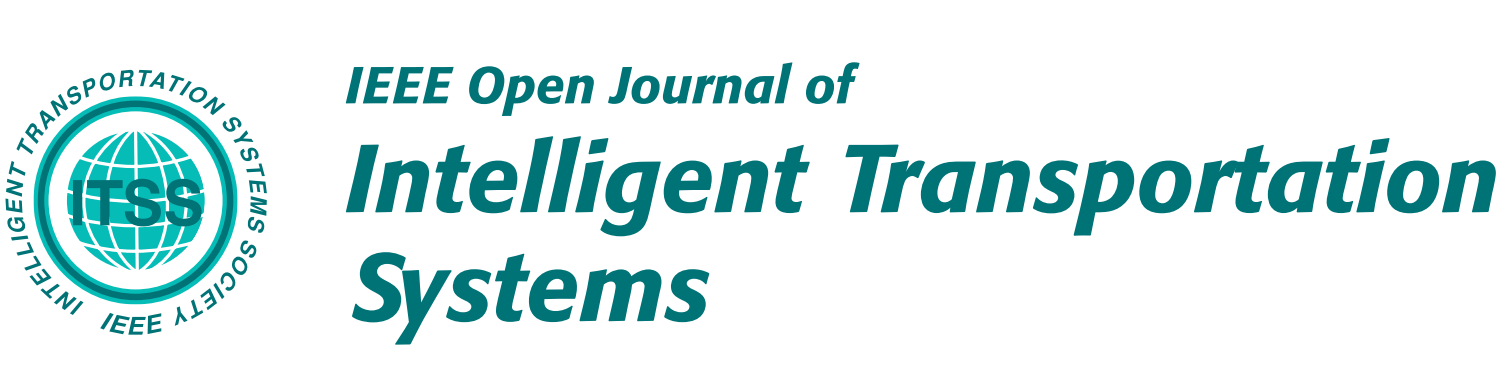}}
\begin{document}
\receiveddate{XX Month, XXXX}
\reviseddate{XX Month, XXXX}
\accepteddate{XX Month, XXXX}
\publisheddate{XX Month, XXXX}
\currentdate{XX Month, XXXX}

\title{Efficient MILP-based Urban Network Traffic Control in Mixed Autonomy with Dynamic Saturation Rates}

\author{Muhammad Haris$^{1}$ \orcidlink{0000-0003-4885-3136}, 
Claudio Roncoli$^{2,1}$ \orcidlink{0000-0002-9381-3021}}

\affil{Department of Built Environment, Aalto University, Finland}
\affil{Mobility and  Industrial Management (MIM), KU Leuven, Belgium}
\corresp{CORRESPONDING AUTHOR: Muhammad Haris (e-mail: muhammad.haris@aalto.fi).}
\authornote{This work was funded by the Research Council of Finland projects AIforLEssAuto (no. 347200) and ALCOSTO (no. 349327).}
\markboth{Mixed Autonomy Traffic Optimization}{Haris \textit{et al.}}

\begin{abstract}
This paper introduces a novel control strategy to optimize urban network traffic in mixed autonomy settings, featuring Connected and Automated Vehicles (CAVs) alongside Human-Driven Vehicles~(HDVs). Unlike previous control strategies, where the impact of driver behavior of CAVs and HDVs is not explicitly considered, we propose a dynamic, queue-responsive saturation flow rate to account for autonomy-driven variations in traffic flow characteristics. The proposed method is based on an extended multi-commodity store-and-forward model to a mixed autonomy environment, integrating optimized routing for CAVs via infrastructure-linked connectivity, and signal timings at every signalized intersection. The problem is formulated as a Non-Convex Quadratic Program (NQP), which accounts for queue evolution, spillback, green time allocation, and CAVs routing. To enable computational efficiency for real-time applications, we transform the NQP into a sequence of convex subproblems, leveraging under- and over-estimators to reformulate it as a Mixed Integer Linear Program (MILP). Experimental results via microscopic simulations validate the efficiency and robustness of the proposed methodology. The results show that the proposed model outperforms the existing multi-commodity approach, thus demonstrating its potential for real-time traffic optimization in future urban mobility systems.
\end{abstract}

\begin{IEEEkeywords}
Mixed autonomy traffic, connected and automated vehicles, store-and-forward modeling, urban network control.
\end{IEEEkeywords}

\maketitle

\section{Introduction}
\IEEEPARstart{U}{rban} traffic congestion poses a significant challenge for cities worldwide, necessitating innovative solutions and effective implementation. One promising avenue to alleviate this issue lies in harnessing emerging technologies, such as connected and automated vehicles (CAVs), alongside advanced traffic management systems \cite{Papamichail2019}. CAVs enable novel control strategies such as platooning, coordinated routing, and dynamic tolling on specific links or paths that can improve mobility without demanding extensive infrastructure upgrades~\cite{Papageorgiou2015}. By leveraging such approaches in a coordinated manner, semi- or fully automated vehicles hold the potential to substantially reduce urban congestion \cite{Foxx2017,Vitale2026}. Moreover, their ability to maintain shorter headways through platooning can boost network capacity, offering a pathway to improved traffic flow in urban settings \cite{Varaiya2017, roncoli2019, lazar2020}.

However, the coexistence of CAVs with HDVs, termed \emph{mixed autonomy}, introduces complexities for traffic management. This mixed traffic composition is expected to persist in urban networks for decades due to gradual adoption of automation, high costs, and regulatory constraints \cite{talebpour2016influence,litman2017autonomous,stead2019automated}. This composition creates uncertainties in traffic behavior and impacts, complicating the design of cohesive management policies. As a result, developing integrated strategies for mixed traffic is increasingly vital, promising significant gains in efficiency, safety, mobility, and sustainability for urban transportation networks \cite{Taiebat2018, Ioannis2020}. Existing traffic management systems, tailored primarily for human-driven vehicles, often focus on individual vehicle benefits and struggle to incorporate the capabilities of automated vehicles, and fail to achieve network-wide optimization.

This research work addresses these challenges by proposing an integrated and coordinated framework for urban networks under mixed traffic conditions, optimizing both traffic signal timings and routing for connected and automated vehicles. We extend the store-and-forward model presented in \cite{desouza2020} to account for the dynamics of mixed traffic. Our approach combines a multi-commodity model for CAVs, guided by routing and signal control, with a single-commodity model for HDVs, focused solely on signal control. A key innovation is the introduction of time- and space-varying saturation rates, reflecting the distinct driving behaviors of these vehicle types. This modeling choice is motivated by the fact that fixed-saturation-rate models assume a constant discharge capacity for each link, whereas in mixed CAV--HDV traffic the effective discharge rate varies with the local time-varying composition of the queue. Since CAVs can maintain shorter headways and benefit from coordination, while HDVs generally maintain longer and more variable headways, the saturation flow rate may change across links and over time. Therefore, using a fixed saturation flow rate would misrepresent the actual discharge capacity when the CAV--HDV composition is heterogeneous. Current multi-commodity traffic control models can represent destination-specific routing decisions, but they do not explicitly account for vehicle-type-dependent discharge behavior. To address this limitation, we introduce a dynamic saturation flow rate that adapts to the mixed autonomy composition of each link. The primary contributions of this work are as follows.
\begin{itemize}
\item{An optimization-based control framework for urban network traffic flow in mixed autonomy, featuring a dynamic saturation flow rate model that captures the time-varying and space-dependent interactions between CAVs and HDVs. Unlike its static counterpart, this dynamic model reflects real-time traffic composition and vehicle coordination levels, particularly those enabled by CAVs. By embedding this dynamic saturation modeling into the control framework, the system more accurately predicts link throughput, leading to significantly improved network efficiency, stability, and responsiveness under varying traffic conditions.}
\item{A convex piecewise Mixed Integer Linear Program (MILP) formulation embedded in a Model Predictive Controller (MPC) framework, is designed to be feasible for real-time implementation.}
\item{An implementation of the proposed MPC as a traffic control strategy in a microscopic simulation with calibrated saturation flow rates.}
\end{itemize}

The paper is organized as follows. Section~\ref{lit_rev} presents the study of related literature. Section~\ref{meth} elaborates on our methodology, detailing the mathematical model, objective function, constraints, and solution approach underpinning the optimization framework. Section~\ref{exp} presents numerical results, describing experimental scenarios and analyzing their outcomes. Finally, Section~\ref{con} concludes the paper with a summary of findings and directions for future research.

\section{Literature Review}
\label{lit_rev}
Recent studies on mixed traffic management have explored both decentralized and centralized strategies to mitigate congestion, enhance throughput, and reduce travel times. This section reviews key methodologies, their contributions, and their limitations, identifying gaps that our research seeks to address.

Decentralized approaches often prioritize local adaptability. For instance,~\cite{zaidi2016back} proposed a multi-commodity framework based on the back-pressure principle~\cite{varaiya2013max}, integrating signal phase activation and adaptive routing in a decoupled manner. By learning neighboring queue states and updating vehicle routes at intersections, their method outperformed single-commodity models. The study in~\cite{liu2018back} extended this work, enhancing its efficacy across larger grid networks. However, these studies do not account for the heterogeneous behavior of HDVs and only partially leverage the unique capabilities of CAVs, such as vehicle-to-infrastructure communication and real-time coordination, which limits their applicability to mixed autonomy scenarios in real-world urban networks. 

Centralized strategies, by contrast, optimize traffic across entire networks. In~\cite{lqmpc2018}, building on earlier work~\cite{le2013linear}, a linear quadratic MPC framework was developed to minimize total network time, using real-time data to align predictions with traffic states. This approach improved flow toward system-optimal levels but struggled with scalability under mixed traffic conditions, where HDVs and CAVs coexist. To address scalability and real-time constraints in large networks, studies such as~\cite{lin2011fast, lin2012efficient} have proposed MPC-based control strategies reformulated as Mixed-Integer Linear Programs (MILPs), which significantly improve computational feasibility without compromising control performance. While effective for macroscopic traffic coordination, these centralized MPC frameworks similarly lack the granularity to differentiate between HDVs and CAVs, and thus do not fully utilize connected vehicle data for adaptive signal control or systematic routing. This gap motivates the development of hybrid or hierarchical control strategies capable of integrating microscopic CAV features within scalable macroscopic optimization models suitable for real-time deployment.

An alternative centralized method in~\cite{li2015hyper, li2017recast} employed a phase-time-traffic hyper-network decomposition, splitting signal control and routing into subproblems solved via Lagrangian relaxations. Their microscopic model, treating each vehicle as an agent with detailed intersection dynamics, reduced travel times but overlooked HDV behavior and the benefits of CAV technologies, such as platooning or guided routing.

Emerging research has begun tackling mixed traffic more explicitly. The article in~\cite{du2022dynamic} introduced a max-pressure-based signal controller that accounts for saturation flow rate variations due to CAVs alongside HDVs. Their findings demonstrated efficiency gains at signalized intersections, even with a simple controller. Similarly,~\cite{lazar2017capacity} explored the impact of platoon formation, developing capacity models based on the Bernoulli principle to show how platooning mitigates shockwaves and boosts flow. More recently, the work in~\cite{Li2026backward} proposed a coordinated platoon--signal control method for signalized intersections in mixed traffic, considering the backward-looking effect. This study is related to mixed traffic control at the intersection and platoon level, while our work focuses on network-level signal timing and CAV routing through a dynamic saturation-rate-based MPC framework.

Furthermore, recent contributions have explored mixed-autonomy traffic control, autonomous-vehicle decision-making, network-wide signal coordination, route recommendation, CAV intersection optimization, and platoon evaluation from complementary perspectives. The work in~\cite{Fatima2025} implemented an attention-based reinforcement learning approach to recover traffic system performance from various disruptions via traffic signal control in mixed-autonomy networks. The study in~\cite{Ye2025spatiotemporal} proposed a deep reinforcement learning method based on spatiotemporal information for network-wide traffic signal coordination control. Similarly,~\cite{Wang2026route} developed a reinforcement-learning-based sequential route recommendation framework for system-optimal traffic assignment. The study in~\cite{Leyer2026} proposed an event-driven decision-making framework for autonomous vehicles in mixed-traffic roundabouts, focusing on vehicle-level decision logic under unsignalized mixed-traffic interactions.

Similarly,~\cite{Vitale2026} proposed a distributed signal-free intersection optimization method for CAVs based on iterative time-slot adjustment and vehicle-level trajectory optimization. The study in~\cite{Hasan2023} provides a simulation-based approach for evaluating platoon safety under communication-related disturbances. These studies highlight the growing importance of mixed-autonomy control, CAV coordination, autonomous-vehicle decision-making, and platooning. However, they do not directly address network-level signal timing and CAV routing with dynamic saturation flow rates under a computationally efficient MPC framework.

To address these aforementioned limitations, our research builds upon foundational work in store-and-forward traffic modeling, including both single-commodity frameworks~\cite{papaStore2009} and their extensions to multi-commodity formulations~\cite{desouza2020}. The multi-commodity approach improves upon the single-commodity paradigm by disaggregating queue lengths by destination, which facilitates destination-specific traffic management, particularly beneficial for routing and signal control involving CAVs. We propose a unified control strategy that integrates signal optimization and routing in mixed autonomy traffic. Our model incorporates dynamic saturation rates, which are endogenously adjusted based on the vehicle composition and behavior on each link. This allows the controller to adapt to real-time variations in CAV--HDV interactions, leading to improved network-level performance and responsiveness.

\begin{table}[htpb] 
\caption{Mathematical Notation} \label{tab:math_notation} \centering \resizebox{\linewidth}{!}{%
\begin{tabular}{@{} l p{0.92\linewidth} @{}} \toprule \textbf{Symbol} & \textbf{Description} \\ \midrule $Z$ & Set of links \\ $J$ & Set of nodes (intersections) \\ $D \subset Z$ & Set of destinations (CAVs and HDVs) \\ $\bar{D}$ & Set of destinations for CAVs, excluding HDVs: $\bar{D} = D - \{0\}$, where $\{0\}$ represents HDVs with no known destination \\ $C$ & Cycle time \\ $K$ & Horizon length in time steps \\ $L_{(j)}$ & Lost time (red phase) per intersection $j$ \\ $T$ & Discrete time step period \\ $k$ & Discrete time index ($k = 0, 1, \ldots, K$) \\ $\tau$ & Control horizon time-window ($\tau = k, \ldots, k+K-1$) \\ $\omega_1,\omega_2,\omega_3,\omega_4$ & Weights in the optimization objective \\ $F_{(z,d)}$ & Cost of the shortest path between link $z$ and destination $d$, calculated by Floyd-Warshall Algorithm \\ $x_{(z,d,k)}$ & Queue length of link $z$ directed to destination $d$ \\ $\tilde{x}_{(z,d,k)}$ & Measured queue length of link $z$ directed to destination $d$ \\ $x_{\max(z)}$ & Maximum queue length of link $z$ \\ $q_{(z,d,k)}$ & Outflow of link $z$ directed to destination $d$ \\ $p_{(z,d,k)}$ & Inflow of link $z$ directed to destination $d$ \\ $b_{(z,d,k)}$ & Demand flow entering link $z$ directed to destination $d$ \\ $r_{(z,d,k)}$ & Exit flow leaving link $z$ directed to destination $d$ \\ $g_{(j,i,k)}$ & Green time at intersection $j$ for phase $i$ \\ $g_{\min(j,i)}$ & Minimum green time at intersection $j$ for phase $i$ \\ $A_{(j)}$ & Set of phases on intersection $j$ \\ $B_{(z)}$ & Set of phases admitting right of way (r.o.w) to link $z$ \\ $O_{(j)}$ & Set of outgoing links at intersection $j$ \\ $I_{(j)}$ & Set of incoming links at intersection $j$ \\ $Y_{(z)}$ & Downstream intersection associated with link $z$ \\ $V_{(z)}$ & Upstream intersection associated with link $z$ \\ $f_{(z,m,d,k)}$ & Transport-flow vector of link $z$ towards downstream link $m$ directed to destination $d$ \\ $G_{(z,m,d,k)}$ & Operational green time of link $z$ towards downstream link $m$ directed to destination $d$ \\ $s_{(z,k)}$ & Dynamic saturation flow rate of link $z$ \\ $h_{\text{HDV}}$ & Average headway of HDVs \\ $h_{\text{CAV}}$ & Average headway of CAVs \\ $\tilde{t}_{\textrm{HDV}(z,m,k)}$ & Observed turning rate of HDVs traversing link $z$ towards link $m$ \\ $t_{\textrm{HDV}(z,m,k)}$ & Turning rate of HDVs traversing link $z$ towards link $m$ \\ $t_{\textrm{CAV}(z,m,d,k)}$ & Turning rate of CAVs traversing link $z$ towards link $m$ towards direction $d$ \\ $e_{(z, k)}$ & Exit turning rate of HDVs at link $z$ \\ $\tilde{e}_{(z,k)}$ & Measured exit turning rate of HDVs at link $z$ \\ $\Theta_{(z,k)}$ & Measure of autonomy level of link $z$ \\ $\varepsilon$ & Very small threshold value \\ $X_{(z,k)}$ & Total queue length of link $z$ \\ $\phi_{(z,k)}$ & Denominator part of the dynamic saturation rate formula \\ $\beta$ & Identical segment of the saturation rate range \\ $\lambda_{(n)}$ & Sequential upper and lower limits for each partition of the saturation rate range \\ $N$ & Number of under- and over-estimator envelopes for MILP \\ $M_{(z,k)}$ & Estimated total number of vehicles in link $z$ \\ $E_{N}$ & Sets of number of envelopes \\ $\omega_{(n,z,k)}$ & Binary variables for the selection of partition $n$ at link $z$ \\ $G_{\max}, G_{\min}$ & Operational green time maximum and minimum value respectively \\ $\phi_{\max}, \phi_{\min}$ & Maximum and minimum value of the denominator part in the saturation rate formula respectively \\ $\Delta s_{(z,k)}$ & Differential part of saturation rate \\ $\Delta \phi_{(n,z,k)}$ & Differential part of the denominator in the saturation rate formula $n$ concerning $\phi_{(z,k)}$ \\ $\Delta X_{(z,k)}$ & Differential part of total queue length \\ \bottomrule \end{tabular}
} 
\par\addvspace{0.4ex} {\raggedright {\footnotesize \textit{Note:} Variables subscripted by $k$ indicate values corresponding to time step $k$.\par}} \end{table}

\section{Methodology}
\label{meth}
This section starts by presenting a mathematical model that builds upon the store-and-forward framework originally developed in \cite{desouza2020}. Our approach unifies two distinct elements: a multi-commodity model tailored for CAVs and a single-commodity model designed for HDVs. The framework assumes that actuation is performed via signal controllers, which regulate both vehicle types, and dynamic routing, which is implemented by CAVs. Furthermore, we exploit the capability of CAVs to sustain reduced headways via defining variable, state-dependent, saturation rates across a multi-destination urban network.

\subsection{Mathematical Model}
\label{math_model}

We model the urban road network as a directed graph comprising a set of arcs (links) $z \in Z$ and nodes (intersections) $j \in J$. Links connect intersections, with specific links designated as entry or exit links, open at one end. Each intersection $j$ is equipped with a controlled traffic signal and is associated with incoming links $i \in I_{(j)}$ and outgoing links $m \in O_{(j)}$. We assume all intersections share a uniform cycle time $C_{(j)} = C, \forall j \in J$, fixed and equal to the model’s discrete time step $T$, i.e., $C = T$, as in \cite{papaStore2009}. The model operates in discrete time, indexed by $k = 0, 1, 2, \ldots, K$, where each step corresponds to a signal cycle of duration $T$, spanning a total time-frame of $KT$.

The signal control plan for each intersection $j$ incorporates a fixed lost time $L_{(j)}$ and consists of a predefined set of phases $A_{(j)}$. The set $B_{(z)}$ denotes phases granting right-of-way (r.o.w.) to link $z$. Green time allocations are determined within this structure and optimized across the network.

Each link’s traffic state is characterized by queue lengths, defined separately for CAVs and HDVs. For CAVs, queue lengths are further segmented by destination, while HDV queues are aggregated without destination specificity. The zeroth link $(z=0)$ represents HDVs as a vehicular type, which are not assigned specific destinations, unlike CAVs. The traffic signal plan, specifically green time durations, is optimized to distribute operational green times across links based on vehicle type and, for CAVs, destination. We assume HDV destinations are unknown, with their distribution to outgoing links governed by observed (or estimated) turning rates. In contrast, CAVs are assigned specific routing commands to their destinations, consistent with \cite{desouza2020}.

To accurately capture the operational nuances of mixed traffic urban networks, we incorporate vehicle-type-dependent behaviors through a dynamic saturation mechanism that evolves with traffic composition. This integration enables a unified control strategy capable of coordinating routing and signal timing under heterogeneous flow conditions. The resulting formulation leads to the following optimization problem: 

\begin{subequations}
\renewcommand{\theequation}{\text{1}.\arabic{equation}}
\begin{flalign}
&\begin{aligned} 
&\text{min} \sum_{k=0}^K\sum_{z\in Z}\sum_{d\in\bar{D}}\frac{x^2_{(z,d,k)}}{x_{\textrm{max(z)}}} + \omega_1\sum_{k=0}^K\sum_{z\in Z}\frac{x^2_{(z,0,k)}}{x_{\textrm{max}(z)}} \\ &+ \omega_2\sum_{z\in Z}\sum_{d\in\bar{D}}F_{(z,d)}x_{(z,d,K)} + \omega_3\sum_{z\in Z}\frac{X^2_{(z, K)}}{x_{\textrm{max}(z)}} \\ &+ \omega_4\sum_{k=1}^K\sum_{j\in J}\sum_{i\in A_{(j)}}\left(g_{(j,i, k)}-g_{(j,i, k-1)}\right)^2 \end{aligned} \label{eq:cost} \\
&\text{s.t.} \nonumber \\
&\begin{aligned}
x_{(z,d,k+1)} &= x_{(z,d,k)} + C\big(p_{(z,d,k)} - q_{(z,d,k)} \\
&+ b_{(z,d,k)} - r_{(z,d,k)}\big), \quad z \in Z, d \in D
\end{aligned} \label{eq:dynamics}
\\
&\begin{aligned}
& p_{(z,d,k)} = \sum_{i \in I_{(j)}} f_{(i,z,d,k)}, \quad z \in Z, d \in D, j = V_{(z)} 
\end{aligned} \label{eq:inflow} \\
& q_{(z,d,k)} = \begin{cases}
\frac{x_{(z,d,k)}}{C}, \quad z = d, d \in D &\\
\sum_{m \in O_{(j)}} f_{(z,m,d,k)}, \quad \parbox{1.5cm}{\raggedright $z\neq d, \newline d \in D, \newline j = Y_{(z)}$}
\end{cases} \label{eq:outflow} \\
& r_{(z,d,k)} = e_{(z)} \frac{x_{(z,d,k)}}{C}, \quad z \in Z \label{eq:exitflow} \\
& f_{(z,m,d,k)} = \begin{cases}
\frac{G_{(z,m,d,k)} s_{(z,k)}}{C}, \quad \parbox{1.8cm}{\raggedright $z \neq d, \newline m \in O_{(j)}, \newline j = Y_{(z)}$} \\
t_{\textrm{HDV}(z,m)} q_{(z,d,k)}, \quad \parbox{1.8cm}{\raggedright $d = 0, \newline m \in O_{(j)}, \newline j = Y_{(z)}$} \\
0 , \quad \parbox{3.5cm}{\raggedright $F_{(z, d)} - F_{(m, d)} \leq \epsilon, \newline m\in O_{(j)}, j=Y_{(z)}$} 
\end{cases} \label{eq:transportflow} \\
& 0 \leq \!\! \sum_{m\in O_{(j)},d\in D} \!\!\! G_{(z,m,d,k)} \leq \sum_{i \in B_{(z)}} \!\! g_{(j,i,k)}, \quad j = Y_{(z)} 
\label{eq:greendist} \\
& \sum_{i\in A(j)} g_{(j,i,k)} = C - L_{(j)}, \quad j \in J  \label{eq:cyclelength} \\
& g_{(j,i,k)} \geq g_{\textrm{min}(j,i)}, \quad i \in A_{(j)}, j \in J \label{eq:mingreen} \\
& X_{(z,k)} = \sum_{d \in \bar{D}} x_{(z,d,k)} + x_{(z,0,k)}, \quad z \in Z \label{eq:queuesum} \\
&0 \leq X_{(z,k)} \leq x_{\textrm{max}(z)}, \quad z \in Z \label{eq:maxqueue} \\
&x_{(z,d,0)} = x_{0(z,d)}, \quad z \in Z, d \in D \label{eq:initial} \\
&s_{(z,k)} = \frac{X_{(z,k)}}{\phi_{(z,k)}} \rightarrow X_{(z,k)} = s_{(z,k)} \phi_{(z,k)}, \quad z \in Z \label{eq:sat} \\
&\phi_{(z,k)} = \!h_\text{CAV} \sum_{d \in \bar{D}} x_{(z,d,k)} + \! h_\text{HDV} x_{(z,0,k)}, \quad z \in Z \label{eq:denom} \\
&\frac{1}{h_\text{HDV}} \leq s_{(z,k)} \leq \frac{1}{h_\text{CAV}}, \quad z \in Z \label{eq:satlimits}
\end{flalign}
\label{eq:optProb}
\end{subequations}

The mathematical notation used in this paper is outlined in Table~\ref{tab:math_notation}. The optimization problem~\eqref{eq:optProb} is a non-convex quadratic program (NQP), which is NP-hard in terms of computational complexity~\cite{vanleeuwen1991}. The various components and modeling choices are explained in the following sections.

\subsection{Objective Function}
\label{obj}
The objective function to be minimized~\eqref{eq:cost} consists of five terms. The first and second terms are the primary components of the optimization problem, aiming to reduce and balance the relative total queue length for each link: the first term pertains to CAVs, while the second term, weighted by $\omega_1$, pertains to HDVs. The first two terms in the objective function are designed in order to reduce the risk of over-saturation and spillback of link queues~\cite{aboudolas2010rolling}. The purpose of the third term, weighted by~$\omega_2$, is to serve as a terminal cost for CAVs, which are also subject to routing decisions. The value~$F_{(z, d)}$ represents the estimated shortest-path distance from each link~$z$ to destination~$d$, which can be computed using a modified version of the Floyd–Warshall algorithm~\cite{floydwarshall}, as detailed in (Appendix~\ref{app:floyd-warshall}). This term captures the expected remaining travel distance for CAVs and encourages the optimization framework to guide them efficiently toward their destinations via the shortest feasible paths. The fourth term, weighted by $\omega_3$, penalizes the composition of larger queues, and it can be adjusted as per the number of vehicular types and CAVs number of destinations. The last term, weighted by $\omega_4$, is a penalty term designed to suppress fluctuations in signal control timing at each intersection $j$ and for each phase $i$ over consecutive time steps; this term ensures that the controller mitigates abrupt shifts in signal timings that would not lead to substantial performance improvements. This formulation effectively constructs the progression of the fourth term in the objective function into a series of distinct, well-defined states.

The weighting parameters present in the objective function~\eqref{eq:cost} can be adjusted, for instance, via a trial-and-error procedure or an adaptive method~\cite{lee2015learning}.
The standard on which comparison can be made is with the trivial setting in which all weights can be zero. Thus, each value of the parameters can be varied by leaving another constant for empirical experimentation in order to get a nominal value for each weighting parameter. One can follow the following suggestions to get the network-configured weights:

\begin{itemize}
    \item Weight $\omega_{1}$ modulates the balance between CAVs in proportion to HDVs: increasing the value of $w_{1}$ prioritizes reducing HDVs queues while decreasing it emphasizes CAV queue management.

    \item The associated weight $\omega_{2}$ in the terminal cost with $\sum_{z\in Z}\sum_{d\in\bar{D}}F_{(z,d)}x_{(z,d,K)}$ incentivizes precise routing of CAVs to their destinations within observed time-frame, reflecting outflow priorities. Increasing $w_{2}$ amplifies this, improving cumulative CAV outflow. Its value is tuned based on the number of CAV destinations in the network.

    \item The weight term $\omega_{3}$ penalizes the total squared queue length across all links within the observed time frame. Increasing its value, the optimization prioritizes minimizing residual queues across all links at the horizon’s end, balancing this objective against earlier queue dynamics. Its magnitude is adjusted relative to the first two terms based on network scale and congestion tolerance.

    \item Signal green timings rapid fluctuations are controlled by the smoothing factor $w_{4}$. An incremental increase in its value dampens oscillations in the controller by the order of $\left[\textrm{sec}^2 \right]$, promoting stability in traffic signal operations. The choice of~$\omega_{4}$ depends on the network layout and the control time scale. It should be selected to provide a good balance between stability and responsiveness, since setting it too high can limit the controller’s ability to react to real-time traffic conditions.
\end{itemize}

\subsection{Constraints}
\label{const}
Constraint~\eqref{eq:dynamics} defines the dynamics of queue lengths for each link as a conservation flow equation, where each link has associated inflows and outflows either from other links denoted by $p$ and $q$, respectively, or from outside sources, denoted by~$b$ and $r$, respectively. Here, the latter part, the sink $r$,  is only defined for HDVs exiting the network. Constraints~\eqref{eq:inflow} and \eqref{eq:outflow} define the inflow~$p_{(z, d, k)}$ and outflow~$q_{(z, d, k)}$ using the transport-flow vector. In constraint~\eqref{eq:outflow}, the first segment ensures that CAVs leave the network upon reaching their respective destinations. Constraint~\eqref{eq:exitflow} defines sinks for HDVs according to the predefined (or measured) exiting turning rates $e_{(z)}$ at each link. 

In constraint~\eqref{eq:transportflow}, a transport-flow vector $f_{(z, m, d, k)}$ is defined to compute the outflow from link $z$ toward link $m$ for both CAVs and HDVs. The first segment uses operational green time with variable saturation rates to determine the respective outflow for CAVs and HDVs, while the second segment dictates that only HDVs $(d=0)$ follow predefined (or measured) turning rates $t_{\textrm{HDV}(z, m)}$, and the third term prevents vehicles to have cyclic looping over the same set of links, i.e., visiting the same links again.

Constraint~\eqref{eq:greendist} allows the controller to allocate intersection time for a particular link $z$ (which has the respective order in right of way $B_{(z)}$) across the outgoing links~$m \in O_{(j)}$ and destinations $d$.  Constraints~\eqref{eq:cyclelength} and \eqref{eq:mingreen} are associated with the controller’s signal timing for each intersection $j$ and the set of phases $i \in A_{(j)}$. Constraint~\eqref{eq:cyclelength} ensures that green times are distributed throughout the entire cycle, while \eqref{eq:mingreen} provides a lower limit for each phase and intersection.

Constraint~\eqref{eq:queuesum} provides the definition of total queue length $X_{(z, k)}$, including both CAVs and HDVs, for each link and time step. Constraint~\eqref{eq:maxqueue} states that queue length is shared among vehicles, i.e., CAVs heading to their intended destinations and HDVs, with a certain limit $x_{\textrm{max}(z)}$ (storage capacity) for each link. Constraint~\eqref{eq:initial} specifies the initial condition for the optimization problem.

Constraint~\eqref{eq:sat} sets the rule for variable saturation rates as a function of queue length and average headways of both CAVs and HDVs defined in the subsequent constraint~\eqref{eq:denom} as denominator $\phi_{(z, k)}$. The denominator $\phi_{(z,k)}$ is bounded by $\phi_{\text{min}} = h_{\text{CAV}} \ x_{\text{min}(z)}$ and $\phi_{\text{max}} = h_{\text{HDV}} \ x_{\text{max}(z)}$, reflecting the range of queue compositions. Finally, constraint~\eqref{eq:satlimits} imposes lower and upper limits on the dynamic saturation rate $s_{(z, k)}$ based on the defined average headways.

Since we assume that saturation flow rates, $s_{(z, k)}$, are dynamic, varying in space, i.e., by link, and in time, we define that the saturation flow rate can be represented by the inverse summation of the average discharging headways \cite{signaltime2015}, as:

\begin{equation}
    s_{(z,k)} = \frac{M_{(z,k)}}{\sum_{i=1}^{M_{(z,k)}} h_i}
    \label{eq:2}
\end{equation}

\begin{figure}[tb]
    \centering
    \includegraphics[scale=0.5]{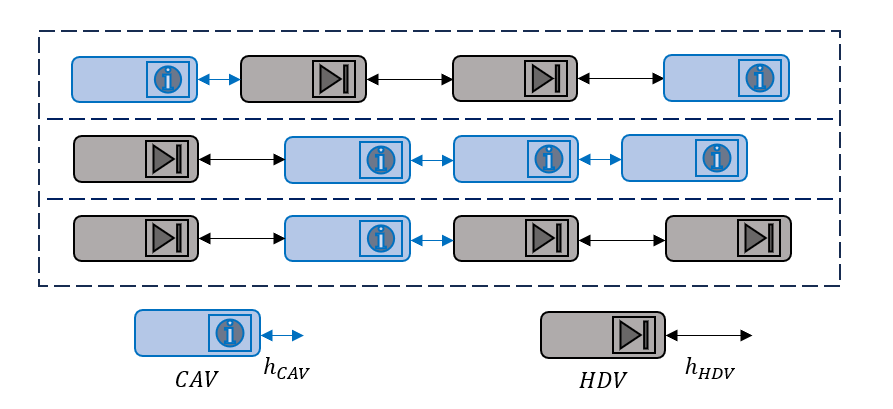}
    \caption{The mixed traffic scenario considered in this paper, where CAVs follow any preceding vehicle maintaining a time-headway~$h_{\textrm{CAV}}$, while HDVs maintain a (potentially longer) time-headway~$h_{\textrm{HDV}}$.}
    \label{fig:1}
\end{figure}

where $h_{i}$ denotes the time-headway of the $i_{th}$ vehicle and $M_{(z,k)}$ is the total number of vehicles in link $z$ at time $k$. We further assume that CAVs can maintain different (usually shorter) time headways with respect to HDVs and can platoon behind any vehicle in front, as sketched in Figure~\ref{fig:1}. Assuming that $h_{\textrm{CAV}}$ and $h_{\textrm{HDV}}$ denote the average time headways of CAVs and HDVs respectively, we employ the mixed traffic autonomy function from~\cite{roncoli2019, lazar2020}, obtaining:

\begin{equation}
    s_{(z, k)} = \frac {1}{\Theta_{(z,k)} \; h_{\textrm{CAV}} + (1-\Theta_{(z,k)}) h_{\textrm{HDV}}} 
\label{eq:3}
\end{equation}

where $\Theta$ defines the autonomy level (percentage of CAVs) of the link or road segment defined as:

\begin{equation}
    \Theta_{(z,k)} = \frac{\sum_{d \in \bar{D}} x_{(z,d,k)}}{X_{(z,k)}} 
\label{eq:4}
\end{equation}

where $X_{(z, k)}$ is the total queue length vector of a link that contains CAVs and HDVs vehicles combined. Thus, by substituting~\eqref{eq:4} in~\eqref{eq:3}, we obtain constraint~\eqref{eq:sat}.

\subsection{Solution Approach}
\label{sol}
To compute the optimal signal timing and CAVs routing, we solve the optimization problem~\eqref{eq:optProb}, assuming the availability of the initial states (queue lengths), the demand vector for both CAVs and HDVs, and the turning and exiting rates for HDVs for the entire optimization horizon; note that the latter can be estimated via methods such as~\cite{lee2014real, ghanim2018estimating} or by using online loop detectors~\cite{shafik2025real}. In our approach, the HDV turning proportions are obtained from observed traffic states and smoothed before being used as input parameters in the MPC controller, as described in Section~\ref{mpc}. Furthermore, we assume that routing information for CAVs is transmitted to each vehicle connected to the system via V2I and that each CAV follows the assigned route.

The optimization problem described in~\eqref{eq:optProb} presents significant computational challenges due to its non-convex structure, stemming primarily from bilinear terms involving products of continuous variables. Specifically, two bilinearities are central to the problem: (i) the \textit{dynamic saturation rates term}, appearing in constraint~\eqref{eq:sat}, and (ii) the \textit{transport-flow term}, present in the first part of constraint ~\eqref{eq:transportflow}. These bilinear expressions result in a non-convex Non-Linear Quadratic Program (NQP), rendering conventional convex optimization techniques ineffective or suboptimal. To address this, we adopt a piecewise relaxation strategy that approximates the non-convex NQP by a Mixed-Integer Quadratic Program (MIQP), and subsequently reformulate it into an MILP through linearization of the quadratic terms in the objective. 

To this end, we develop a piecewise MILP-quadratic relaxation approach, which improves tractability while maintaining a tight approximation of the original problem. The method builds on classical McCormick under- and over-estimators and leverages disjunctive programming techniques for formulating piecewise linear approximations~\cite{wicaksono2008}. The variable of interest is partitioned into $N$ uniform segments, each bounded by a pair of envelope constraints. Two reformulation schemes are commonly used to represent such disjunctions: the Big-M formulation and the convex hull formulation. We employ the latter due to its superior tightness and numerical stability, especially when dealing with bilinear terms.

We apply the convex hull-based reformulation to the dynamic saturation rate $s_{(z,k)}$ as defined in constraint~\eqref{eq:sat}, enabling a piecewise-linear convex outer approximation of the bilinear term $X_{(z,k)} = s_{(z,k)} \phi_{(z,k)}$. To ensure regularity and avoid numerical instability, we use uniform partitioning (rather than arbitrary or adaptive segmentation) across the domain of $s_{(z,k)}$, which is characterized as:

\begin{subequations}
\renewcommand{\theequation}{\text{5}.\arabic{equation}}
\begin{align}
&\begin{aligned} \beta = \frac{(1/h_\text{CAV} - 1/h_\text{HDV})}{N} = \frac{h_\text{HDV} - h_\text{CAV}}{N(h_\text{HDV} \ h_\text{CAV})} \end{aligned} \\
&\begin{aligned} \lambda_{(n)} = 1/h_\text{HDV} \!+\! n \beta , \quad n \!\in \! E_{N+1} \!=\! \{1, 2, ...,\! N\!+\!1\} \end{aligned}
\end{align}
\label{eq:sat_segmentation}
\end{subequations} 
where $N$ is the number of envelopes of under- and over-estimators for the piecewise MILP relaxation, $\beta$ is the identical segment length for variable $s_{(z, k)}$, and $\lambda_{(n)}$ are the sequential upper and lower bounds for each partition $n$. 

Next, we introduce binary variables to indicate the active segment:
\begin{subequations} 
\renewcommand{\theequation}{\text{6}.\arabic{equation}}
\begin{align}
&\begin{aligned}\omega_{(n, z, k)}=\begin{cases}
1, \ \lambda_{(n)} \!\leq\! s_{(z, k)} \!\leq\! \lambda_{(n+1)}, \quad n \in E_{N} 
\\
0, \quad \text{otherwise}
\end{cases} \end{aligned} \\
&\begin{aligned} \sum_{n \in E_{N}} \omega_{(n, z, k)} = 1, \quad z \in Z \end{aligned}
\end{align} 
\label{eq:active_set}
\end{subequations}

The bilinear term $ X_{(z, k)} = s_{(z, k)} \phi_{(z, k)} $ can be recast into a convex approximation, as follows: 

\begin{subequations} 
\renewcommand{\theequation}{\text{7}.\arabic{equation}}
\begin{flalign}
&\begin{aligned} 
s_{(z, k)} &= (1/h_\text{HDV}) + \beta \sum_{n \in E_{N}} (n - 1) \ \omega_{(n, z, k)} \\
&+ \Delta s_{(z, k)}, \quad z \in Z
\end{aligned} \\[2ex]
&0 \leq \Delta s_{(z, k)} \leq \beta, \quad z \in Z
\end{flalign}
\begin{flalign}
&\phi_{(z, k)} = \phi_{min} + \sum_{n \in E_{N}} \Delta \phi_{(n, z, k)}, \quad z \in Z\\
&\begin{aligned} 
0 \leq \Delta \phi_{(n, z, k)} \leq ( \phi_{max} - \phi_{min} ) \ \omega_{(n, z, k)}, & \\ \ z \in Z, n \in E_{N}
\end{aligned} \\[2ex]
&\Delta X_{(z, k)} \leq ( \phi_{max} - \phi_{min} ) \ \Delta s_{(z, k)}, \quad z \in Z \\
&\Delta X_{(z, k)} \leq \beta \ ( \phi_{(z, k)} - \phi_{min}), \quad z \in Z \\
&\begin{aligned} 
\Delta X_{(z, k)} &\geq ( \phi_{max} - \phi_{min} ) \ \Delta s_{(z, k)} \\ & + \beta \ (\phi_{(z, k)} - \phi_{max}), \quad z \in Z 
\end{aligned} \\[2ex]
&\begin{aligned} 
X_{(z, k)} &= (1/h_\text{HDV})\phi_{(z, k)} + \beta \sum_{n \in E_{N}} (n-1) \ \Delta \phi_{(n, z, k)} \\ 
&+ \Delta X_{(z, k)}, \quad z \in Z
\end{aligned} \\[2ex]
&\phi_{min} \leq \phi_{(z, k)} \leq \phi_{max}, \quad z \in Z
\label{eq:sat_convex}
\end{flalign}
\end{subequations} 

The bilinear term of transport-flow vector, i.e. $f_{(z, m, d, k)} = G_{(z, m, d, k)} \ (s_{(z, k)} / C) $, can be approximated by McCormick relaxation technique over its known bounds, as follows:

\begin{subequations} 
\renewcommand{\theequation}{\text{8}.\arabic{equation}}
\begin{flalign}
&\begin{aligned} 
f_{(z, m, d, k)} &\geq \frac{1}{C} \left( s_{(z, k)} G_{\text{min}} + \frac{1}{h_\text{HDV}}(G_{(z, m, d, k)}-G_{\text{min}}) \right), \\
&z \in Z, d \in D, m \in O_{(j)}, j=Y_{(z)}
\end{aligned} \\
&\begin{aligned}
f_{(z, m, d, k)} &\geq \frac{1}{C} \left( s_{(z, k)} G_{\text{max}} + \frac{1}{h_\text{CAV}}(G_{(z, m, d, k)}-G_{\text{max}}) \right), \\
& z \in Z, d \in D, m \in O_{(j)}, j=Y_{(z)}
\end{aligned} \\ \nonumber
&\begin{aligned}
f_{(z, m, d, k)} &\leq \frac{1}{C} \left( s_{(z, k)} G_{\text{max}} + \frac{1}{h_\text{HDV}}(G_{(z, m, d, k)}-G_{\text{max}}) \right), \\
& z \in Z, d \in D, m \in O_{(j)}, j=Y_{(z)}
\end{aligned} \\
\end{flalign}
\begin{flalign}
&\begin{aligned}
f_{(z, m, d, k)} &\leq \frac{1}{C} \left( s_{(z, k)} G_{\text{min}} + \frac{1}{h_\text{CAV}}(G_{(z, m, d, k)}-G_{\text{min}}) \right), \\
& z \in Z, d \in D, m \in O_{(j)}, j=Y_{(z)}
\end{aligned}
\end{flalign}
\label{eq:transportflow_convex}
\end{subequations} 

The complete MIQP problem is formed by substituting constraints ~\eqref{eq:transportflow}-first part and~\eqref{eq:sat} with the relaxed forms~\eqref{eq:transportflow_convex} and~\eqref{eq:sat_convex}, respectively, and adding binary constraint~\eqref{eq:active_set}. To further enhance solvability, the quadratic terms in the objective function (i.e., first, second, and fourth terms) are approximated using piecewise linearization, as suggested in~\cite{fourer1985simplex}, thereby converting the MIQP into an MILP. This final formulation enables efficient solution using commercial MILP solvers while preserving a high-fidelity approximation of the original non-convex problem.

\subsection{Integration with Model Predictive Control}
\label{mpc}
To enable real-time adaptability of the proposed optimization framework, we integrate an MPC scheme~\cite{camacho2016mpc}, formulated over a receding horizon. The optimization problem at each decision step \(k\) is defined using the piecewise MILP-relaxed formulation of the original non-convex program~\eqref{eq:optProb}, as constructed in equations~\eqref{eq:sat_convex} and \eqref{eq:transportflow_convex}. We assume the control horizon matches the prediction horizon in steps (number of cycles). The MPC controller optimizes decision variables over a finite prediction horizon \(K\), including the signal timings \( g_{(j,i,k)} \), disaggregated effective green times \( G_{(z, m, d, k)} \), which allow to calculate CAVs routing variables \( t_{\textrm{CAV}(z,m,d,k)} \) as:

\begin{equation} 
t_{\textrm{CAV} (z, m, d, k)} = \frac{G_{(z, m, d, k)}}{\displaystyle\sum_{\substack{m \in O_{(j)} \\ j = Y_{(z)}}} G_{(z, m, d, k)}}.
\label{eq:cav_turn}
\end{equation}

Only the control actions determined for the first step are implemented, while the horizon window is shifted forward in the next iteration to account for updated traffic conditions.

Real-time system states such as link-specific queue lengths \( \tilde{x}_{(z,d, k)} \), HDV turning proportions \( \tilde{t}_{\textrm{HDV}(z,m, k)} \), and exit turnings \( \tilde{e}_{(z, k)} \), are incorporated as feedback to the MPC controller. This feedback mechanism enables dynamic adjustment of the control strategy based on observed urban traffic states, modeled through store-and-forward dynamics that approximate the macroscopic evolution of queues and flows across the network. To handle the erratic nature of HDV turning behaviors in urban networks, the turning proportions \( \tilde{t}_{\textrm{HDV}(z,m, k)} \) are smoothed using an exponential weighted moving average, ensuring stable inputs to the MPC while maintaining responsiveness to recent observations. The smoothed HDV turning proportion at step \(k\) as:

\begin{equation} 
t_{\textrm{HDV}(z,m,k)} =
\alpha \tilde{t}_{\textrm{HDV}(z,m,k)}
+ (1-\alpha)t_{\textrm{HDV}(z,m,k-1)},
\label{eq:hdv_smoothing}
\end{equation}

where $\tilde{t}_{\textrm{HDV}(z,m,k)}$ is the observed turning proportion, and~$\alpha=0.9$ is selected to prioritize recent observations in the MPC update. This value provides a responsive estimate for time-varying HDV turning behavior while retaining a small contribution from the previous estimate to reduce sensitivity to instantaneous measurement noise.

\begin{figure}[tb]
    \centering    
    \includegraphics[width=\columnwidth]{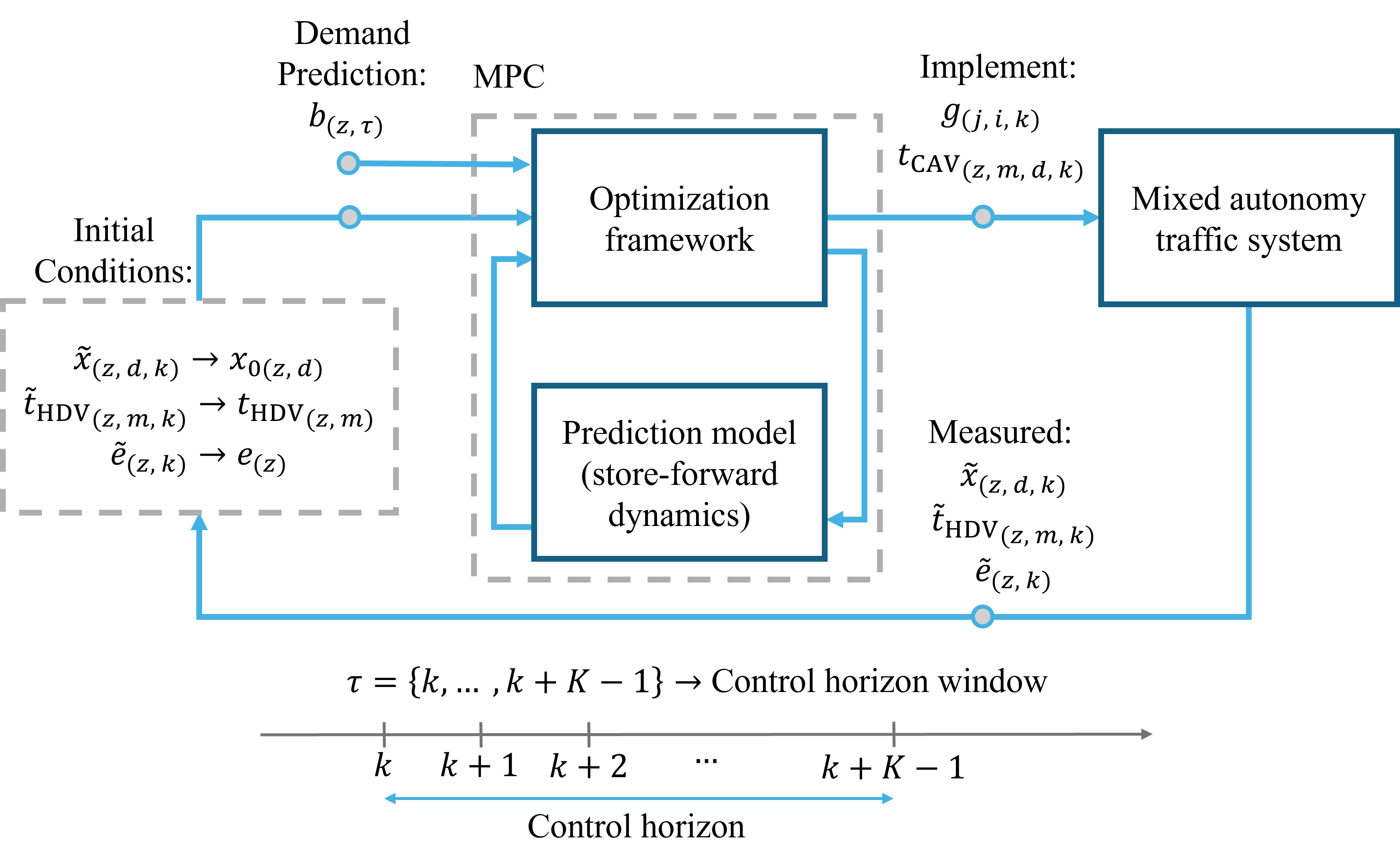}
    \caption{The MPC framework employed in this paper. At time step \( k \), the solver optimizes over horizon \( K \) using input states along with the store-and-forward model dynamics, applying only the first control step before shifting the horizon to the next steps.}
    \label{fig:mpc}
\end{figure}

This closed-loop control architecture ensures responsiveness to stochastic and time-varying phenomena such as fluctuating demand and uncertain HDV turning behaviors. Although validated through simulation, the design is structured to operate with real-time traffic data, making it amenable to practical deployment. The schematic overview of the MPC framework is illustrated in Figure~\ref{fig:mpc}.

\section{Experimental Setup}
\label{exp}
To validate the effectiveness of the proposed control framework under mixed autonomy conditions, we evaluate its performance within a simulated urban network using the microscopic traffic simulator Aimsun Next~\cite{aimsun}. The control logic is implemented externally and iteratively interacts with the simulation environment via the Aimsun Next API at each simulation step.

\begin{figure}[tb]
\centering
\begin{tikzpicture}
[
    node distance=1.5cm, 
    >=Stealth, 
    intersection/.style={rectangle, draw, thick, minimum size=0.4cm, inner sep=0pt, color = black}, 
    entry/.style={coordinate, draw, fill=green!30, minimum size=4mm}, 
    exit/.style={coordinate, draw, fill=red!30, minimum size=4mm}, 
    link/.style={->, thick, teal}, 
    label distance=1mm, 
    every node/.append style={font=\footnotesize} 
]
    \node[intersection] (J1) at (0, 0) {$1$};
    \node[intersection] (J2) at (1.2, 0) {$2$};
    \node[intersection] (J3) at (2.4, 0) {$3$};
    \node[intersection] (J4) at (3.6, 0) {$4$};
    \node[intersection] (J5) at (0, -1.2) {$5$};
    \node[intersection] (J6) at (1.2, -1.2) {$6$};
    \node[intersection] (J7) at (2.4, -1.2) {$7$};
    \node[intersection] (J8) at (3.6, -1.2) {$8$};
    \node[intersection] (J9) at (0, -2.4) {$9$};
    \node[intersection] (J10) at (1.2, -2.4) {$10$};
    \node[intersection] (J11) at (2.4, -2.4) {$11$};
    \node[intersection] (J12) at (3.6, -2.4) {$12$};
    \node[intersection] (J13) at (0, -3.6) {$13$};
    \node[intersection] (J14) at (1.2, -3.6) {$14$};
    \node[intersection] (J15) at (2.4, -3.6) {$15$};
    \node[intersection] (J16) at (3.6, -3.6) {$16$};

    \draw[link] (J1) -- (J2) node[midway, above] {$3$};
    \draw[link] (J2) -- (J3) node[midway, above] {$6$};
    \draw[link] (J3) -- (J4) node[midway, above] {$9$};
    \draw[link] (J8) -- (J7) node[midway, above] {$19$};
    \draw[link] (J7) -- (J6) node[midway, above] {$17$};
    \draw[link] (J6) -- (J5) node[midway, above] {$15$};
    \draw[link] (J9) -- (J10) node[midway, above] {$24$};
    \draw[link] (J10) -- (J11) node[midway, above] {$26$};
    \draw[link] (J11) -- (J12) node[midway, above] {$28$};
    \draw[link] (J16) -- (J15) node[midway, above] {$37$};
    \draw[link] (J15) -- (J14) node[midway, above] {$35$};
    \draw[link] (J14) -- (J13) node[midway, above] {$33$};

    \draw[link] (J1) -- (J5) node[midway, left] {$4$};
    \draw[link] (J6) -- (J2) node[midway, left] {$7$};
    \draw[link] (J3) -- (J7) node[midway, left] {$10$};
    \draw[link] (J8) -- (J4) node[midway, left] {$13$};
    \draw[link] (J5) -- (J9) node[midway, left] {$16$};
    \draw[link] (J10) -- (J6) node[midway, left] {$18$};
    \draw[link] (J7) -- (J11) node[midway, left] {$20$};
    \draw[link] (J12) -- (J8) node[midway, left] {$22$};
    \draw[link] (J9) -- (J13) node[midway, left] {$25$};
    \draw[link] (J14) -- (J10) node[midway, left] {$27$};
    \draw[link] (J11) -- (J15) node[midway, left] {$29$};
    \draw[link] (J16) -- (J12) node[midway, left] {$31$};

    \node[entry] at (-1.2, 0) (E1) {$E_1$};
    \draw[link] (E1) -- (J1) node[midway, above] {$1$};
    \node[entry] at (-1.2, -1.2) (E2) {$E_2$};
    \draw[link] (J5) -- (E2) node[midway, above] {$14$};
    \node[entry] at (-1.2, -2.4) (E3) {$E_3$};
    \draw[link] (E3) -- (J9) node[midway, above] {$23$};
    \node[entry] at (-1.2, -3.6) (E4) {$E_4$};
    \draw[link] (J13) -- (E4) node[midway, above] {$32$};
    \node[entry] at (0, 1.2) (E5) {$E_5$};
    \draw[link] (E5) -- (J1) node[midway, left] {$2$};
    \node[entry] at (1.2, 1.2) (E6) {$E_6$};
    \draw[link] (J2) -- (E6) node[midway, left] {$5$};
    \node[entry] at (2.4, 1.2) (E7) {$E_7$};
    \draw[link] (E7) -- (J3) node[midway, left] {$8$};
    \node[entry] at (3.6, 1.2) (E8) {$E_8$};
    \draw[link] (J4) -- (E8) node[midway, left] {$11$};

    \node[exit] at (4.8, 0) (X1) {$X_1$};
    \draw[link] (J4) -- (X1) node[midway, above] {$12$};
    \node[exit] at (4.8, -1.2) (X2) {$X_2$};
    \draw[link] (X2) -- (J8) node[midway, above] {$21$};
    \node[exit] at (4.8, -2.4) (X3) {$X_3$};
    \draw[link] (J12) -- (X3) node[midway, above] {$30$};
    \node[exit] at (4.8, -3.6) (X4) {$X_4$};
    \draw[link] (X4) -- (J16) node[midway, above] {$39$};
    \node[exit] at (0, -4.8) (X5) {$X_5$};
    \draw[link] (J13) -- (X5) node[midway, left] {$34$};
    \node[exit] at (1.2, -4.8) (X6) {$X_6$};
    \draw[link] (X6) -- (J14) node[midway, left] {$36$};
    \node[exit] at (2.4, -4.8) (X7) {$X_7$};
    \draw[link] (J15) -- (X7) node[midway, left] {$38$};
    \node[exit] at (3.6, -4.8) (X8) {$X_8$};
    \draw[link] (X8) -- (J16) node[midway, left] {$40$};
\end{tikzpicture}
\caption{Grid network with \( |Z| = 40 \) links and \( |J| = 16 \) intersections used for testing the proposed methodology.}
\label{fig:net}
\end{figure}
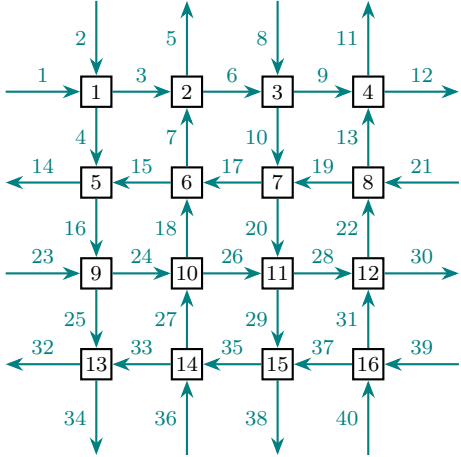

The considered urban network consists of \( |Z| = 40 \) links and \( |J| = 16 \) signalized intersections, including 8~entry links, 8 exit links, and 24 interlinks, as shown in Figure~\ref{fig:net}. Signalized intersections are operated using fixed green time phases assigned according to link orientation (vertical and horizontal). Each signal cycle has a duration of \( C = 120 \, \text{s} \), and the simulation spans a total of \( 30 \) cycles (steps), corresponding to a one-hour duration. Within this simulation, the MPC controller operates over a prediction horizon \( K \), solving a receding horizon optimization problem at each control step (equivalent to the cycle).

For each intersection \( j \in J \), a lost time of \( L_{(j)} = 10 \, \text{s} \) is uniformly allocated between two signal phases. Green times are constrained by a minimum of \( g_{\min(j,i)} = 30 \, \text{s} \) for each phase \( i \in A_{(j)} \), and queue lengths are capped at a maximum of \( x_{\max(z)} = 40 \, \text{veh} \) for all links \( z \in Z \).

Additionally, in the simulation, all the intersections are signalized, each featuring two incoming and two outgoing links. Every link of the urban network is a single-lane road with a length of $l = 200 \ \text{m}$. Each incoming link is controlled by a single stage: the first stage grants right-of-way to the horizontal links (as depicted in Figure~\ref{fig:net}), while the second stage provides right-of-way to the vertical links.

Traffic demand (veh/h), including origin-destination (OD) pairs for CAVs and only origins (0) for HDVs, is illustrated in Table~\ref{tab:dem}. 

The proposed control strategy is implemented in Python, interfacing with AMPL~\cite{fourer1990modeling} to formulate the optimization problem, which is solved using Fico-Xpress~\cite{fico_xpress}.

\begin{table*}[tb]
\centering
\caption{Demand data (veh/h) for CAV\MakeLowercase{s} and HDV\MakeLowercase{s} across time intervals}
\label{tab:dem}
\begin{tabular}{l c c c c c c}
\toprule
\textbf{Vehicular Type} & \textbf{Origin link} & \textbf{Destination link} & \textbf{0--20 (min)} & \textbf{20--40 (min)} & \textbf{40--50 (min)} & \textbf{50--60 (min)} \\ \toprule
\multirow{13}{*}{CAVs}
& 1 & 12 & 216 & 288 & 288 & 0\\
& 2 & 34 & 36 & 72 & 72 & 0\\
& 8 & 38 & 216 & 288 & 288 & 0\\
& 21 & 14 & 144 & 180 & 144 & 0\\
& 23 & 30 & 72 & 108 & 144 & 0\\
& 36 & 5 & 180 & 216 & 216 & 0\\
& 39 & 32 & 108 & 144 & 144 & 0\\
& 40 & 11 & 216 & 288 & 288 & 0\\
& 39 & 5 & 108 & 144 & 144 & 0\\
& 2 & 38 & 180 & 216 & 216 & 0\\
& 23 & 11 & 144 & 180 & 144 & 0\\
& 21 & 32 & 72 & 108 & 144 & 0\\
& 36 & 12 & 36 & 72 & 72 & 0\\
\midrule
\multirow{8}{*}{HDVs}
& 1 & 0 & 216 & 288 & 288 & 0\\
& 2 & 0 & 216 & 288 & 288 & 0\\
& 8 & 0 & 216 & 288 & 288 & 0\\
& 21 & 0 & 216 & 288 & 288 & 0\\
& 23 & 0 & 216 & 288 & 288 & 0\\
& 36 & 0 & 216 & 288 & 288 & 0\\
& 39 & 0 & 216 & 288 & 288 & 0\\
& 40 & 0 & 216 & 288 & 288 & 0\\
 \bottomrule
    \vspace{0pt}
\end{tabular}
\\
\footnotesize{Note: The zeroth link (0) denotes random exiting links for HDVs, chosen stochastically by the simulator.}
\end{table*}

\subsection{Simulation Configuration}
\label{simulation}
In the simulation, CAVs are modeled using the Cooperative Adaptive Cruise Control (CACC) module as their car-following behavior, as described in~\cite{milanes2014modeling}. HDVs adhere to the Gipps model, the default car-following model in Aimsun Next~\cite{gipps1981behavioural}. For traffic assignment, HDVs follow a binomial stochastic route choice model to capture their probabilistic turning behavior at intersections. In contrast, CAVs follow optimized routing decisions determined by the designed MPC controller at each intersection. This centralized control assumes all vehicles maintain a communication link with the traffic infrastructure, enabling real-time information exchange. In the absence of MPC control, CAVs default to the shortest-path routing provided by Aimsun Next, which is also used when the control logic is deactivated or fixed-green signals.

The microscopic car-following parameters for the Gipps and CACC models are kept at the default Aimsun Next settings throughout all replications, while the parameters directly used in the proposed macroscopic controller are the calibrated average discharge headways of HDVs and CAVs.

Each experiment is repeated for ten replications consisting of randomized seed values to simulate driver behavior stochasticity.
When referring to multiple replications, results are reported as the median of efficiency metrics over all the replications. In particular, the following metrics are computed. The Median Percent Error (MPE) is defined as:

\begin{equation}
\text{MPE} = \text{median} \left( \frac{s_{\text{model},i} - s_{\text{sim},i}}{s_{\text{sim},i}} \times 100 \right) (\%),
\label{eq:mpe}
\end{equation}

where \( s_{\text{model},i} \) and \( s_{\text{sim},i} \) represent the modeled and simulated saturation flow rates, respectively, for the \( i \)-th observation i.e. link and time-step pair.
The Median Absolute Deviation (MAD) is given by:

\begin{equation}
\text{MAD} = \text{median} \left( | s_{\text{model},i} - s_{\text{sim},i} | \right) (\textrm{veh/h}),
\label{eq:mad}
\end{equation}

where \( | s_{\text{model},i} - s_{\text{sim},i} | \) denotes the absolute difference between the modeled and simulated saturation flow rates for the \( i \)-th observation. For each individual link (in a replication), the MPE and MAD are computed by aggregating over all time intervals of that link. For the entire network, the MPE and MAD are computed by aggregating over all links and all time intervals within the given replication.

\subsection{Activation Logic Design}
\label{act_logic}
In an actual implementation, deploying a control strategy that alters routing and green times may be unnecessary when there are low network-wide queues, as congestion is very unlikely to occur. To overcome this issue, an activation logic can be deployed so that the control strategy is activated only when it is deemed necessary~\cite{papamichail2010,iordanidou2015,tajdari2020feedback}. To this purpose, we define:

\begin{equation}
\Gamma_\textrm{control}(k) = 
\begin{cases} 
1, \quad \text{if } \max_{z} X_{(z, k)} > X_\textrm{act} \\ 
0, \quad \text{if } \max_{z} X_{(z, k)} < X_\textrm{deact} \\ 
\Gamma_\textrm{control}(k-1), \quad \text{otherwise,}
\end{cases}
\label{eq:act_logic}
\end{equation}

where \(X_\textrm{act}\) and \(X_\textrm{deact}\), with $ X_\textrm{act} > X_\textrm{deact}$, are the activation and deactivation queue thresholds, respectively. As a result, the controller activates when the maximum of the network-wide queue length exceeds \(X_\textrm{act}\) and remains active until it drops below \(X_\textrm{deact}\). To prevent frequent switching between states due to fluctuations in traffic flow, the activation logic preserves the previous state when the maximum network-wide queue length stays in between the two states \(X_\textrm{deact} < \max_{z} \  X_{(z, k)} < X_\textrm{act}\). When the control strategy is deactivated, CAVs receive routing information from the controller to prevent misrouting or vehicles deviating to unintended destinations, as dynamically changing routes mid-operation may lead to coordination failures.

\subsection{Controlled vs. Uncontrolled Strategies}
In our microscopic experiments, three strategies are compared: 1)~the proposed mixed autonomy MPC-controlled strategy with dynamic saturation flow rate (DynamicSF), 2)~ a modified version of the multi-commodity MPC-controlled strategy with fixed constant saturation flow rate~\cite{desouza2020} (ConstantSF), and 3)~an uncontrolled baseline scheme (FixedTime) with fixed signal timings of \( g_{(j,i,k)} = 55~\text{s} \) per phase, with each phase consisted of lost time of $L_{(j)} = 5~\text{s}$, and no centralized routing applied to CAVs. In this case, CAVs follow the default shortest path routing handled by Aimsun Next, similar to HDVs. The ConstantSF scenario models mixed traffic with a fixed saturation flow rate, determined using the known demand proportion CAV--HDV alongside their respective average headways derived from Aimsun Next simulations. Specifically, with an average headway of 1.8 seconds for CAVs and 2.7 seconds for HDVs, and a 50-50 percent CAV--HDV demand proportion, the saturation flow rate for the ConstantSF scenario is calculated using Equation~\eqref{eq:3} and~\eqref{eq:4}: $ s_{(z, k)} = 1600 ~\text{veh/h}, ~ \forall z \in Z, k \in K $. Thus, this scenario assumes a uniform distribution of CAVs and HDVs in the network. The weights for both controlled strategies are equivalently set to compare the results; $w_{1} = 1, w_{2} = 10, w_{3} = 100$ and $w_{4} = 0.001 $. The weights are adjusted by trial-and-error as per the considered scenario i.e. network configuration, input traffic demand, control actions, and overall observed network flow. 

\section{Numerical Results}
\label{res}
This section presents the outcomes of the microscopic simulation experiments conducted in Aimsun Next to evaluate the proposed mixed autonomy solution of DynamicSF against ConstantSF, and the unoptimized FixedTime.

\subsection{Queue Dynamics \& Saturation Flow Analysis}
This subsection evaluates the DynamicSF model, highlighting its ability to capture saturation flow rates that adapt to varying queue lengths.

\subsubsection{Network Error Metrics} 

\begin{table}[!t]
\caption{Network-wide performance metrics of DynamicSF across multiple replications for varying values with $N$}
\centering
\begin{tabular}{c c c c c c}
\toprule
\multirow{2}{*}{\textbf{Replication No.}} & \multicolumn{2}{c}{\textbf{$N=5$}} & \multicolumn{2}{c}{\textbf{$N=7$}} \\
\cmidrule(lr){2-3} \cmidrule(lr){4-5}
& \textbf{MPE} & \textbf{MAD} & \textbf{MPE} & \textbf{MAD} \\
& (\%) & (veh/h) & (\%) & (veh/h) \\
\toprule
1	&	0.12	&	60.67	&	0.34	&	60.31	\\
2	&	0.46	&	60.84	&	0.47	&	55.37	\\
3	&	0.34	&	63.57	&	-0.16	&	63.29	\\
4	&	0.45	&	58.55	&	-0.03	&	65.68	\\
5	&	-0.12	&	58.25	&	-0.03	&	58.95	\\
6	&	0.11	&	54.92	&	-0.03	&	58.78	\\
7	&	0.01	&	53.82	&	-0.03	&	59.02	\\
8	&	0.05	&	55.93	&	0.26	&	59.57	\\
9	&	0.06	&	62.23	&	0.20	&	68.12	\\
10	&	0.04	&	60.50	&	-0.33	&	63.35	\\
\bottomrule
\vspace{0.0pt}
\end{tabular}
\label{tab:network_metrics}
\end{table}
\begin{table}[!t]
  \centering
\caption{Link-specific error metrics of DynamicSF for network saturation flow rates in replication no. 5 with $N=7$}
\begin{tabular}{c c c c c}
    \toprule
    \textbf{Link No.} & \textbf{MPE} & \textbf{MAD} & \textbf{M-Mod.} & \textbf{M-Sim.} \\
    & (\%) & (veh/h) & (veh/h) & (veh/h) \\
\toprule
   1 &       0.33 &      20.19 &    1555.79 &    1560.23 \\
   2 &      -1.23 &      37.01 &    1624.49 &    1599.76 \\
   3 &      -1.41 &      37.49 &    1526.69 &    1510.83 \\
   4 &       6.42 &     135.96 &    1512.99 &    1724.89 \\
   6 &      -0.03 &      93.43 &    1568.79 &    1510.83 \\
   7 &      -6.44 &     137.13 &    1667.81 &    1527.58 \\
   8 &       2.74 &      66.89 &    1576.68 &    1627.09 \\
   9 &      -0.72 &      41.98 &    1581.18 &    1543.47 \\ 
  10 &       3.46 &      79.57 &    1644.14 &    1656.71 \\
  13 &       2.73 &     105.03 &    1728.47 &    1764.55 \\
  15 &      -2.75 &      71.55 &    1582.31 &    1543.16 \\
  16 &       0.99 &      77.90 &    1638.88 &    1646.84 \\
  17 &      -0.12 &      28.54 &    1681.06 &    1642.19 \\
  18 &       6.32 &     153.32 &    1564.78 &    1629.40 \\
  19 &      -0.01 &      29.30 &    1691.00 &    1653.50 \\
  20 &       2.91 &      96.91 &    1577.36 &    1598.80 \\
  21 &       0.36 &      13.63 &    1578.34 &    1603.89 \\
  22 &       0.04 &      92.14 &    1741.67 &    1714.10 \\
  23 &      -1.35 &      35.52 &    1596.89 &    1595.68 \\
  24 &       0.29 &      34.20 &    1647.02 &    1632.47 \\
  25 &      -0.19 &      66.38 &    1552.38 &    1555.30 \\
  26 &      -0.69 &      45.32 &    1715.91 &    1682.48 \\
  27 &      -0.70 &      48.63 &    1611.38 &    1632.43 \\
  28 &      -2.69 &      73.42 &    1669.18 &    1582.36 \\
  29 &      -1.74 &     127.20 &    1707.97 &    1629.40 \\
  31 &       2.20 &      75.65 &    1668.21 &    1675.23 \\
  33 &       1.68 &      56.78 &    1517.29 &    1499.72 \\
  35 &      -0.02 &      81.67 &    1506.74 &    1499.72 \\
  36 &      -2.03 &      46.28 &    1609.67 &    1592.86 \\
  37 &      -0.51 &      39.31 &    1476.59 &    1477.29 \\
  39 &      -1.51 &      34.24 &    1614.64 &    1590.39 \\
  40 &      -0.72 &      42.78 &    1596.99 &    1599.76 \\
\bottomrule
\vspace{0.0pt}
\end{tabular} 
\label{tab:error_gap_metrics}
\end{table}

The simulations, comprising 10 replications, are conducted using the demand presented in Table~\ref{tab:dem}, with a prediction horizon $K = 3$ and MILP approximation envelopes of $N = 5$ and $N = 7$. Table~\ref{tab:network_metrics} summarizes network-wide performance, capturing saturation flow rate variations between modeled and simulated values across 10 replications, evaluated using MPE and MAD, elaborated mathematically in~\eqref{eq:mpe} and~\eqref{eq:mad}, respectively. In contrast, Table~\ref{tab:error_gap_metrics} presents robust error metrics for all links in the network for replication No. 5, based on the same MPE and MAD criteria, additionally reporting median values of modeled and simulated saturation flow rates. 

From Table \ref{tab:network_metrics}, it is evident that the MPE across all replications remains at or below $0.47\%$, reflecting a strong correspondence between modeled and simulated saturation flow rates. Additionally, the MAD does not exceed $68$ veh/h, suggesting low variability despite the presence of outliers. The consistency in median saturation flow rates between the model and simulation further underscores the robust performance of the DynamicSF model. 

While increasing the number of MILP envelopes to $N = 7$ yields a marginal improvement over the $N = 5$ configuration reflected in slightly reduced MPE and MAD, the enhancement remains limited. This can be attributed to the inherent gap between theoretical optimization models and the practical implementation constraints of microscopic traffic simulation.
In particular, the MILP-based solution assumes idealized flow control and continuous flow response to signal timings. However, microscopic simulators such as Aimsun Next incorporate more granular and realistic driver behavior, vehicle dynamics, stochastic arrival patterns, and minimum headway enforcement. These aspects introduce nonlinearities and delays that cannot be fully captured by the piecewise-linear approximations used in the MILP formulation, even with a higher resolution (i.e., greater $N$).

Moreover, saturation flow in simulation is also influenced by factors such as reaction time, vehicle acceleration profiles, and queuing dynamics, which are approximated in the macroscopic model used in our problem formulation. Therefore, although the optimization output becomes mathematically more accurate with finer envelope granularity, its practical translation into realized saturation flows remains constrained by simulation realism.

\subsubsection{Queue Length Evolution vs Saturation Flow Rates Variation}
\begin{figure*}[tb]
    \captionsetup{
        font=small, 
        labelfont=up, 
        textfont=up, 
        justification=centering}
    \captionsetup[subfloat]{
        font=small, 
        labelfont=up, 
        textfont=up,
        position=bottom}
    \centering
    \subfloat[\label{fig:quesat3}]{\includegraphics[width=0.39\textwidth,keepaspectratio]{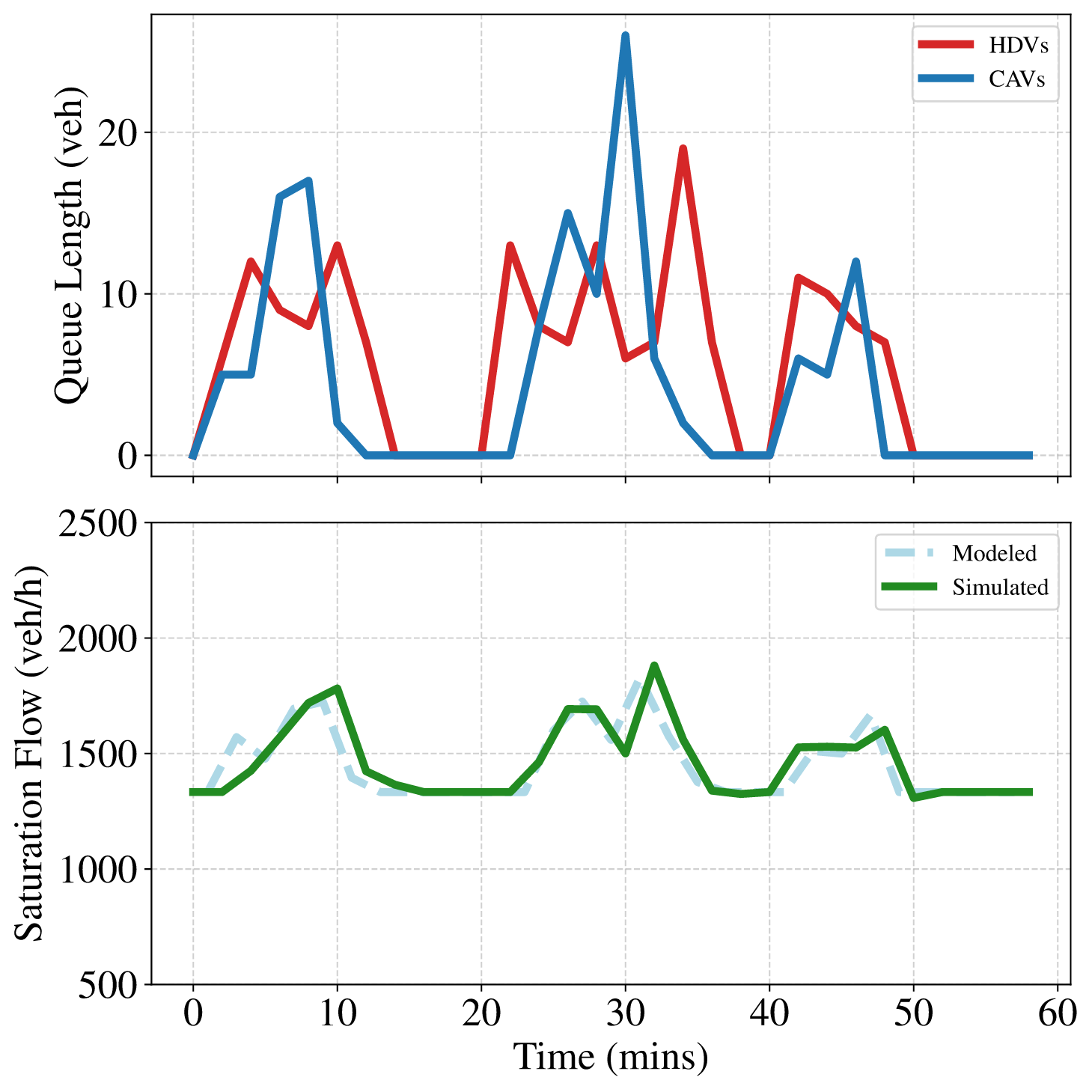}}\hspace{0.5cm}
    \subfloat[\label{fig:quesat9}]{\includegraphics[width=0.39\textwidth,keepaspectratio]{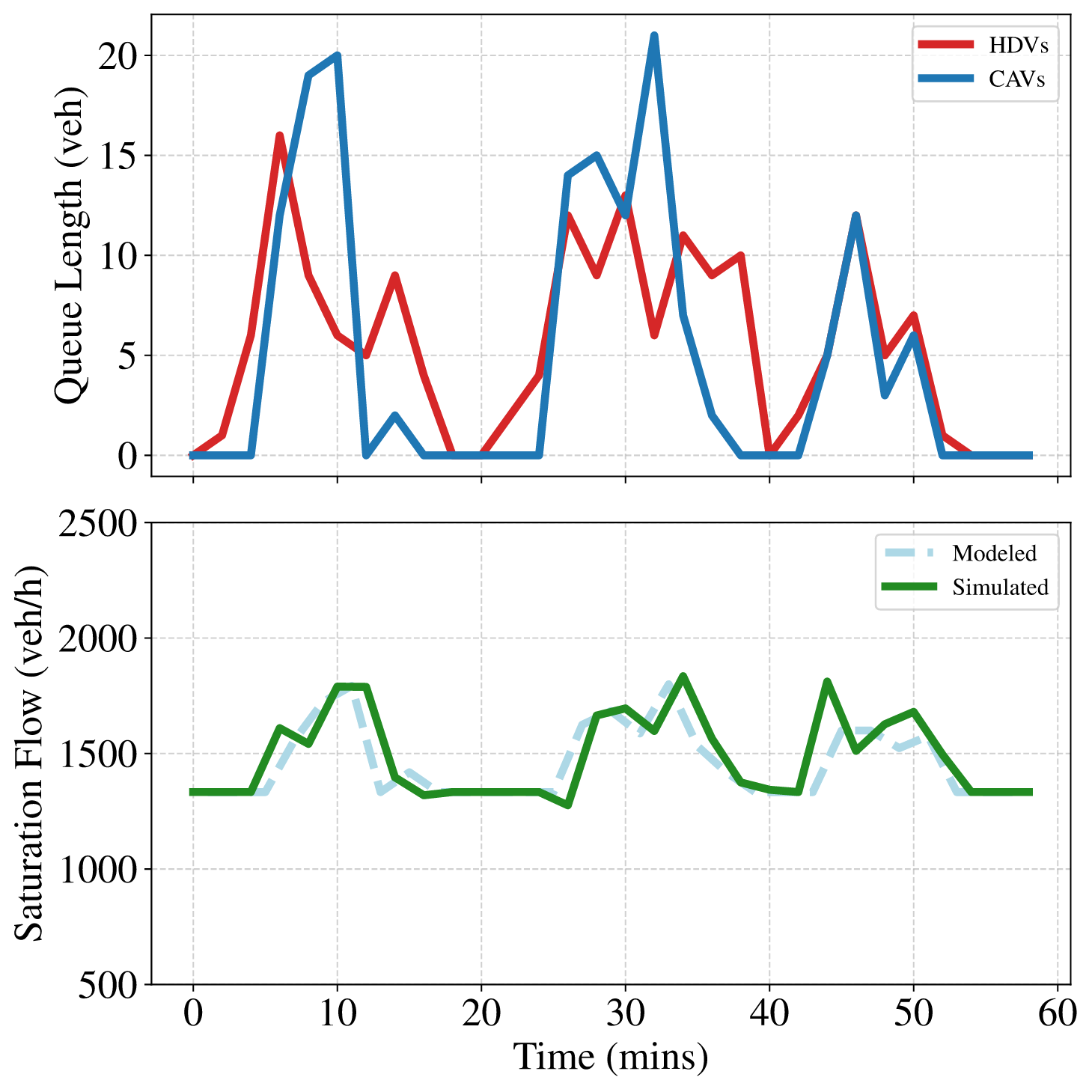}}
    \vspace{0.1cm}
    
    \subfloat[\label{fig:quesat17}]{\includegraphics[width=0.39\textwidth,keepaspectratio]{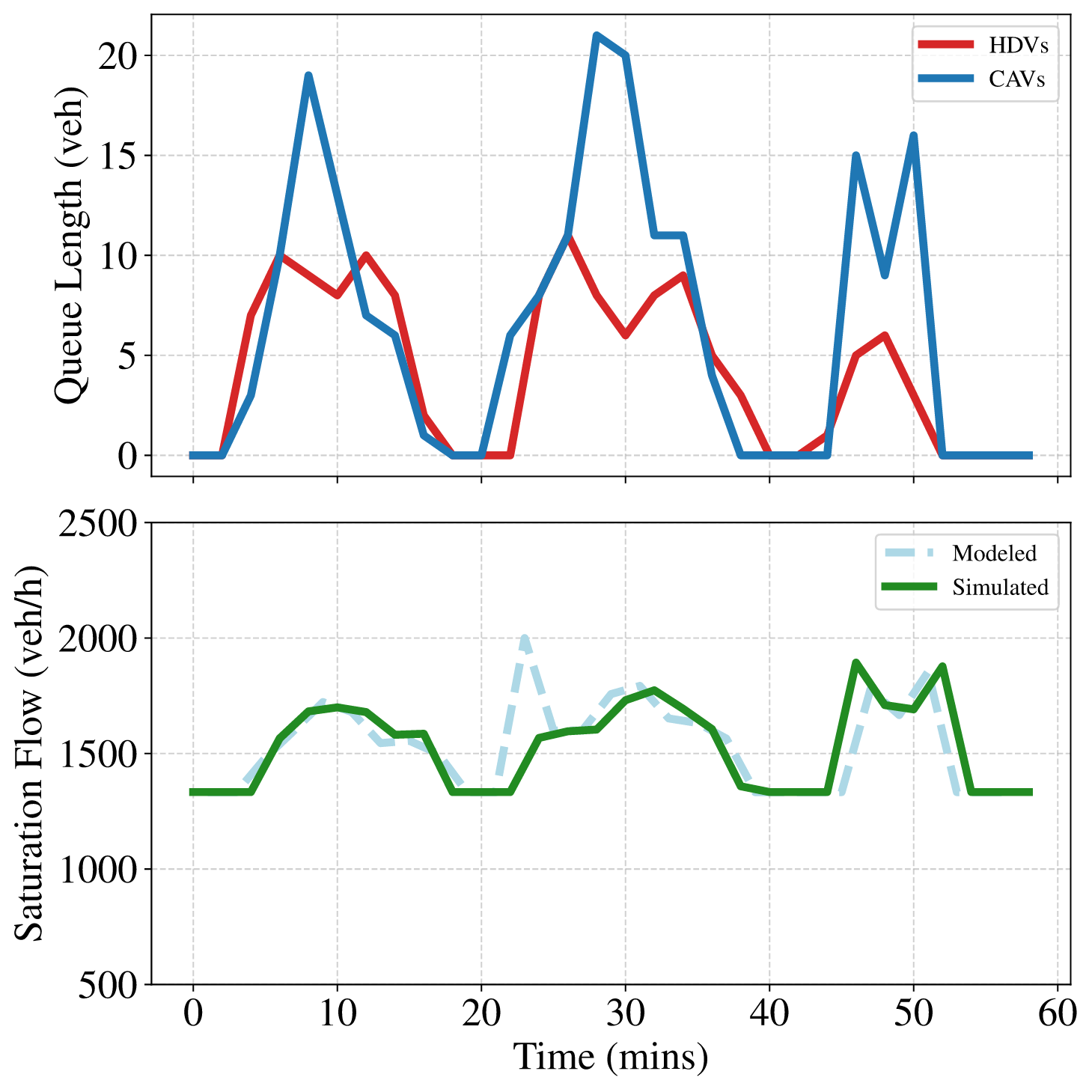}}\hspace{0.5cm}
    \subfloat[\label{fig:quesat26}]{\includegraphics[width=0.39\textwidth,keepaspectratio]{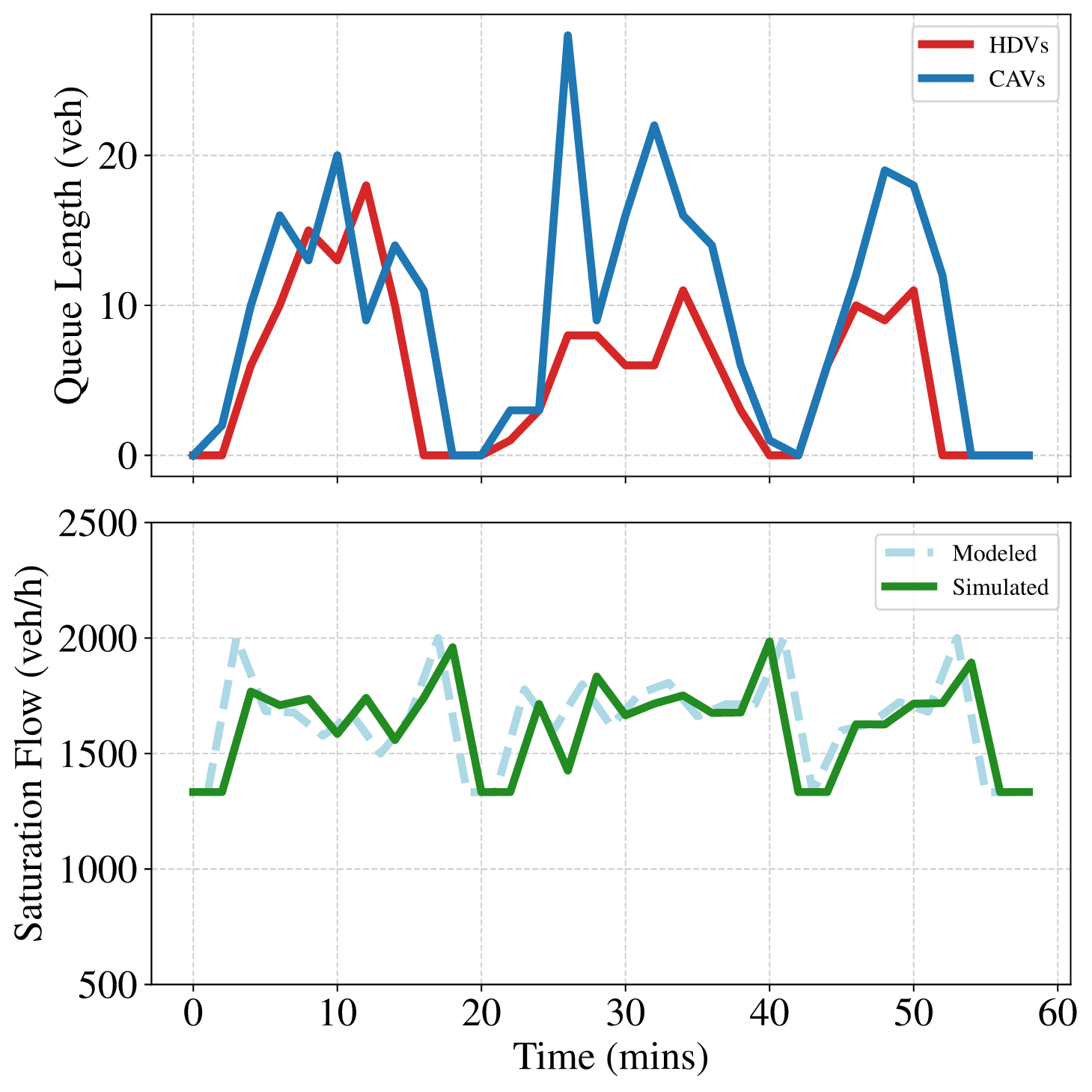}}
    \caption{Evolution of queue lengths (veh) and varying saturation flow rates (veh/h) for: (a) Link 3, (b) Link 9, (c) Link 17, and (d) Link 26.}
    \label{fig:queue_evolution}
\end{figure*}

A series of figures illustrating the relationship between saturation flow rates and queue lengths for a representative set of arterial links in the network, specifically links $\{3, 9, 17, 26\}$, are presented in Figure~\ref{fig:queue_evolution}. These illustrations, depicting saturation flow rates alongside corresponding queue lengths for the specified links, are derived from Replication No. $5$ (which represents the median results) , as detailed in Table \ref{tab:error_gap_metrics}.
The dynamic saturation rate exhibits a range of variation between $1333.3$ veh/h and $2000$ veh/h, effectively adapting to the queue composition comprising CAVs and HDVs. This adaptability highlights the robustness of the model in accommodating diverse traffic scenarios (different distribution). Notably, the modeled simulation values demonstrate a fine alignment with the simulated values which are the reflection of changing headways of different vehicular types under consideration. 


\subsection{Signal Control Performance}
\label{sec:sign_perf}
\begin{figure}[tb]
  \centering
\includegraphics[width=0.8\columnwidth]{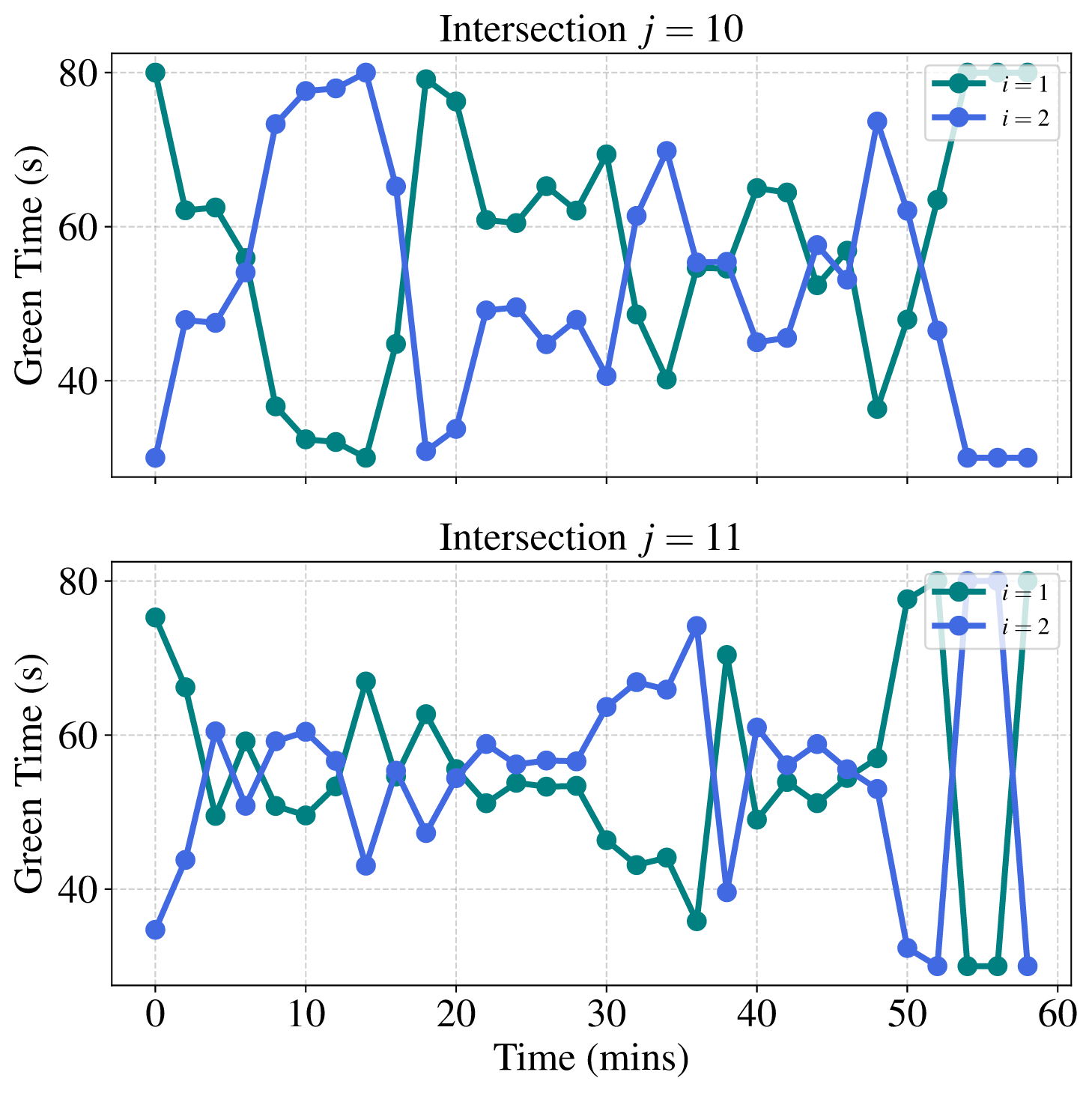}
  \caption{Green signal timing profiles for smoothing factor $w_4 = 0.0001$, illustrating abrupt transitions across controller phases.}
  \label{fig:signal1}
\end{figure}

\begin{figure}[tb]
  \centering
\includegraphics[width=0.8\columnwidth]{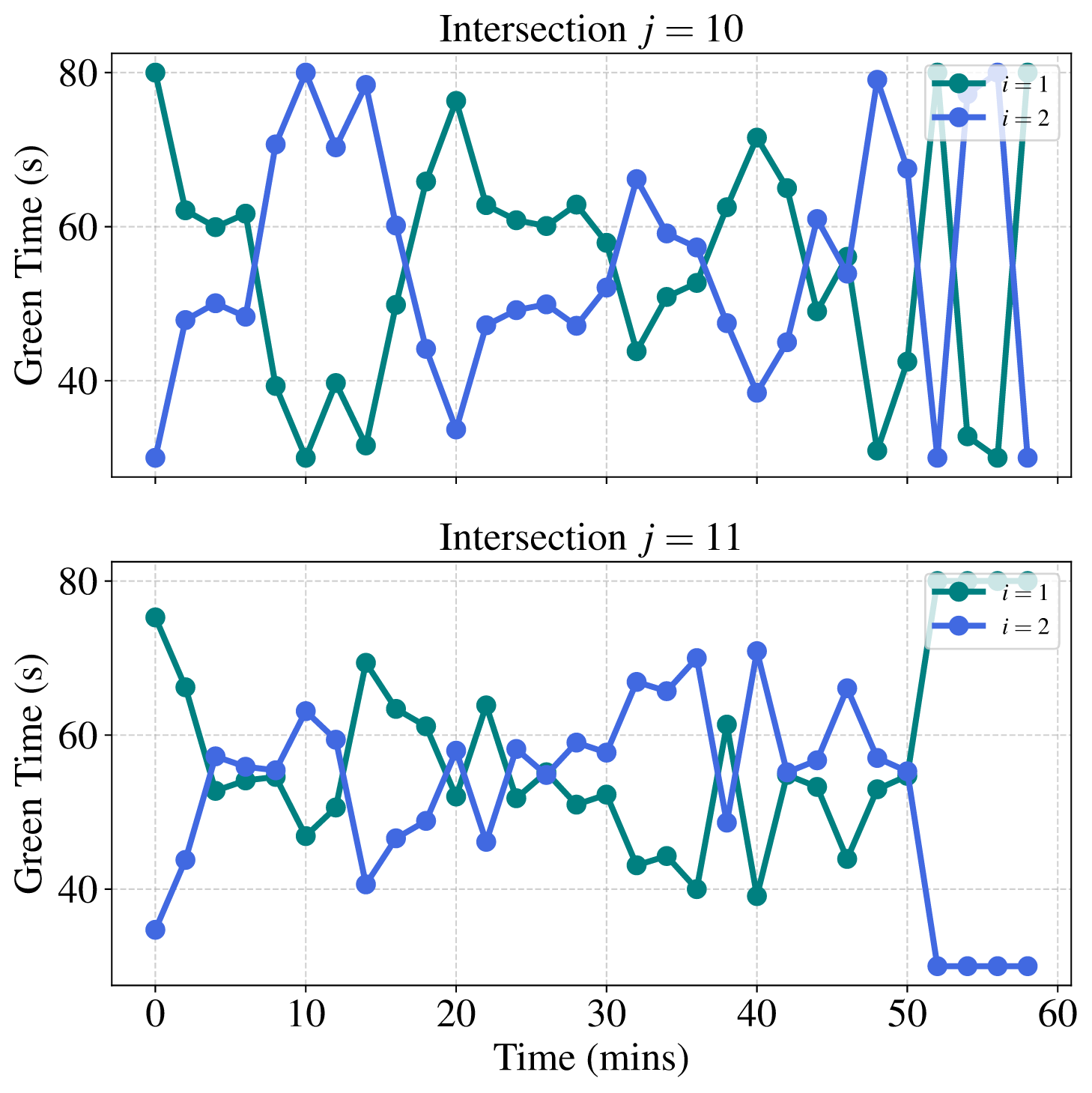}
  \caption{Green signal timing profiles for smoothing factor $w_4 = 0.001$, demonstrating smoother variations across controller phases.}
  \label{fig:signal2}
\end{figure}

This subsection examines signal timing adaptations under the DynamicSF model for selected intersections $ \{10, 11\}$. All intersections are divided into two phases, i.e. $i = 1 \  \text{and} \ i = 2$ for horizontal flow and vertical flow, respectively, varying between 30 s and 80 s over the simulation. 
The signal performance is assessed by comparing two distinct smoothing factor weights, $w_4 = 0.0001$ and $w_4 = 0.001$, as outlined in Section~\ref{obj}. A comparative analysis of the results, illustrated in Figures~\ref{fig:signal1}--\ref{fig:signal2}, reveals notable differences in the green time profiles. Specifically, the green signal timings exhibit more abrupt transitions when $w_4 = 0.0001$ compared to the smoother variations observed with $w_4 = 0.001$. This suggests that increasing the smoothing factor weight can effectively reduce abruptness in the green signal timing across all phases of the controller. However, achieving this enhanced smoothness incurs an additional optimization cost, reflecting a trade-off between signal stability and computational complexity.


\subsection{Efficiency Traffic Metrics}
\label{efftraffmet}
Efficiency metrics compare DynamicSF, ConstantSF, and FixedTime across prediction horizon variations (\( K = 2, 3, 5 \)) considering the following performance indicators:
\begin{itemize}
\item Total Mean Queue (TMQ): The average queue length of the network consisted of all the links;
\item Aggregate Travel Time (ATT): Total travel time experienced per vehicle to traverse the network;
\item Delay: Average delay time per vehicle per kilometer for the whole network.
\end{itemize}
The activation logic is implemented in all the tested scenarios of DynamicSF and ConstantSF, as described in Section~\ref{act_logic}. The thresholds are set as \( X_\textrm{act} = 20 \)~veh and \( X_\textrm{deact} = 10 \)~veh, based on preliminary simulations identifying critical congestion levels. To ensure consistency in evaluating the DynamicSF strategy, the number of MILP approximation envelopes is fixed at $N = 5$ across all experiments.

\begin{table*}[tb] 
\centering
  \caption{Comparison between DynamicSF, ConstantSF, and FixedTime}
  \label{tab:eff_traffic}
  \begin{tabular}{c *{7}{c}}
    \toprule
    \textbf{Strategy} & \textbf{Prediction Horizon} 
    & \multicolumn{2}{c}{\textbf{TMQ (veh)}} 
    & \multicolumn{2}{c}{\textbf{ATT (h)}} 
    & \multicolumn{2}{c}{\textbf{Delay (s/km)}} \\
    \cmidrule(lr){3-4} \cmidrule(lr){5-6} \cmidrule(lr){7-8}
     & & \centering \textbf{Avg} & \centering \textbf{Std dev} 
       & \centering \textbf{Avg} & \centering \textbf{Std dev} 
       & \centering \textbf{Avg} & \centering \textbf{Std dev} \tabularnewline
    \toprule
    \centering FixedTime      & \centering -- & \centering 158 & \centering 7.0   & \centering 340 & \centering 6.1   & \centering 141 & \centering 5.7   \tabularnewline
    \centering ConstantSF     & \centering 2  & \centering 134 & \centering 5.8 & \centering 310 & \centering 4.7 & \centering 118 & \centering 4.2   \tabularnewline
    \centering DynamicSF      & \centering 2  & \centering \textbf{119}  & \centering 3.7   & \centering \textbf{298} & \centering 4.5   & \centering \textbf{106} & \centering 3.7   \tabularnewline
    \centering ConstantSF     & \centering 3  & \centering 116  & \centering 7.5   & \centering 298 & \centering 6.8   & \centering 100  & \centering 7.1   \tabularnewline
    \centering DynamicSF      & \centering 3  & \centering \textbf{115} & \centering 3.3   & \centering \textbf{297} & \centering 3.1   & \centering \textbf{98} & \centering 2.4
   \tabularnewline
    \centering ConstantSF     & \centering 5  & \centering 105
  & \centering 7.8   & \centering 286  & \centering 7.0  & \centering 93  & \centering 6.6 \tabularnewline
    \centering DynamicSF      & \centering 5  & \centering \textbf{99}  & \centering 2.7   & \centering \textbf{280} & \centering 2.7   & \centering \textbf{87}  & \centering 1.9   \tabularnewline
    \bottomrule
    \vspace{0.0pt}
  \end{tabular} 
\end{table*}

Results presented in Table \ref{tab:eff_traffic} demonstrate that the DynamicSF outperforms all the other strategies. The DynamicSF is better than ConstantSF for shorter or larger horizon lengths, though both strategies show lower TMQ, ATT, and Delays than the FixedTime solution. The choice of prediction horizon changes the efficiency of the MPC-based strategy. 

The FixedTime strategy serves as a baseline, exhibiting a TMQ of $158$ veh, ATT of $340$ h, and Delay of $141$ s/km. In contrast, both ConstantSF and DynamicSF substantially improve these metrics across all horizons. Notably, DynamicSF outperforms ConstantSF in most cases, achieving lower ATT (e.g., $280$ h vs. $286$ h for $K = 5$) and Delay (e.g., $87$ s/km vs. $93$ s/km at $K = 5$), with consistently lower standard deviations, implying improved travel time reliability. For~TMQ, DynamicSF excels at a higher horizon ($99$ veh vs. $105$ veh at $K = 5$), while at lower horizon ~$= 2$ ($119$ veh vs. $134$ veh). The impact of the prediction horizon reveals that DynamicSF performs best at $K = 5$ but degrades slightly with shorter horizons, while ConstantSF shows mixed trends, improving from horizon length of $K = 3$ to $K = 5$ and performing worst at $K = 2$ for ATT and Delay. 

\begin{figure}[tb]
  \centering
\includegraphics[width=0.9\columnwidth]{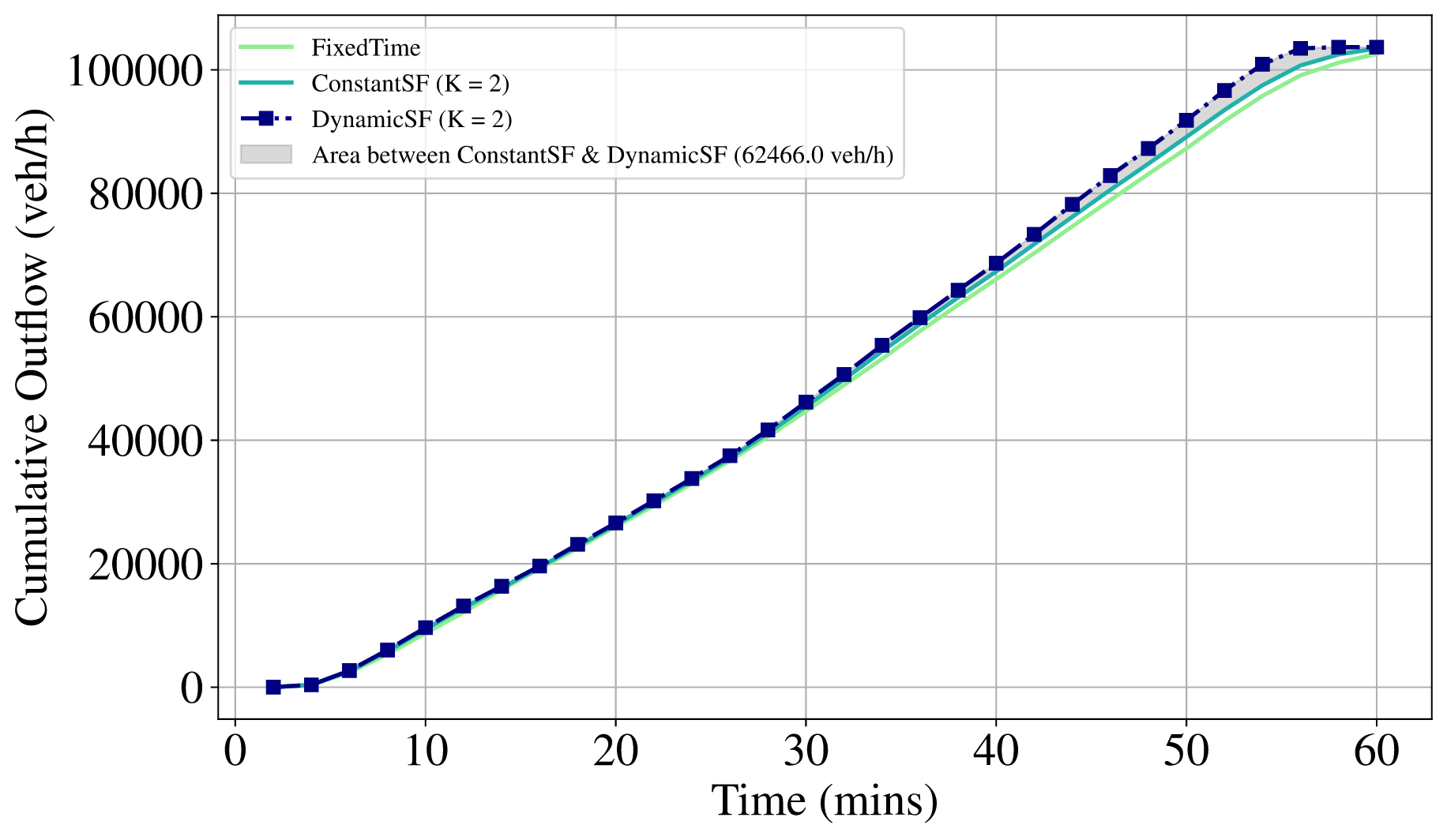}
  \caption{Network-wise cumulative outflow (veh/h) comparison across strategies FixedTime, ConstantSF, and DynamicSF.}
\label{fig:cum_flow}
\end{figure}

Additionally, the cumulative outflow analysis between all the strategies, i.e. FixedTime, DynamicSF ($K = 2$) and ConstantSF ($K = 2$), is depicted in Figure~\ref{fig:cum_flow}. The traffic state reaches gridlock condition around $t = 30$ mins. The ConstantSF cumulative outflow curve is performing slightly better than FixedTime. The DynamicSF strategy maintains a higher outflow rate, enabling all vehicles to exit the network more efficiently compared to ConstantSF as well as FixedTime. 
The shaded area between the DynamicSF and ConstantSF cumulative outflow curves illustrates the enhanced performance of DynamicSF over ConstantSF, highlighting its slightly superior performance in sustaining higher vehicle outflow throughout the simulation.

\begin{figure}[tb]
  \centering
\includegraphics[width=0.9\columnwidth]{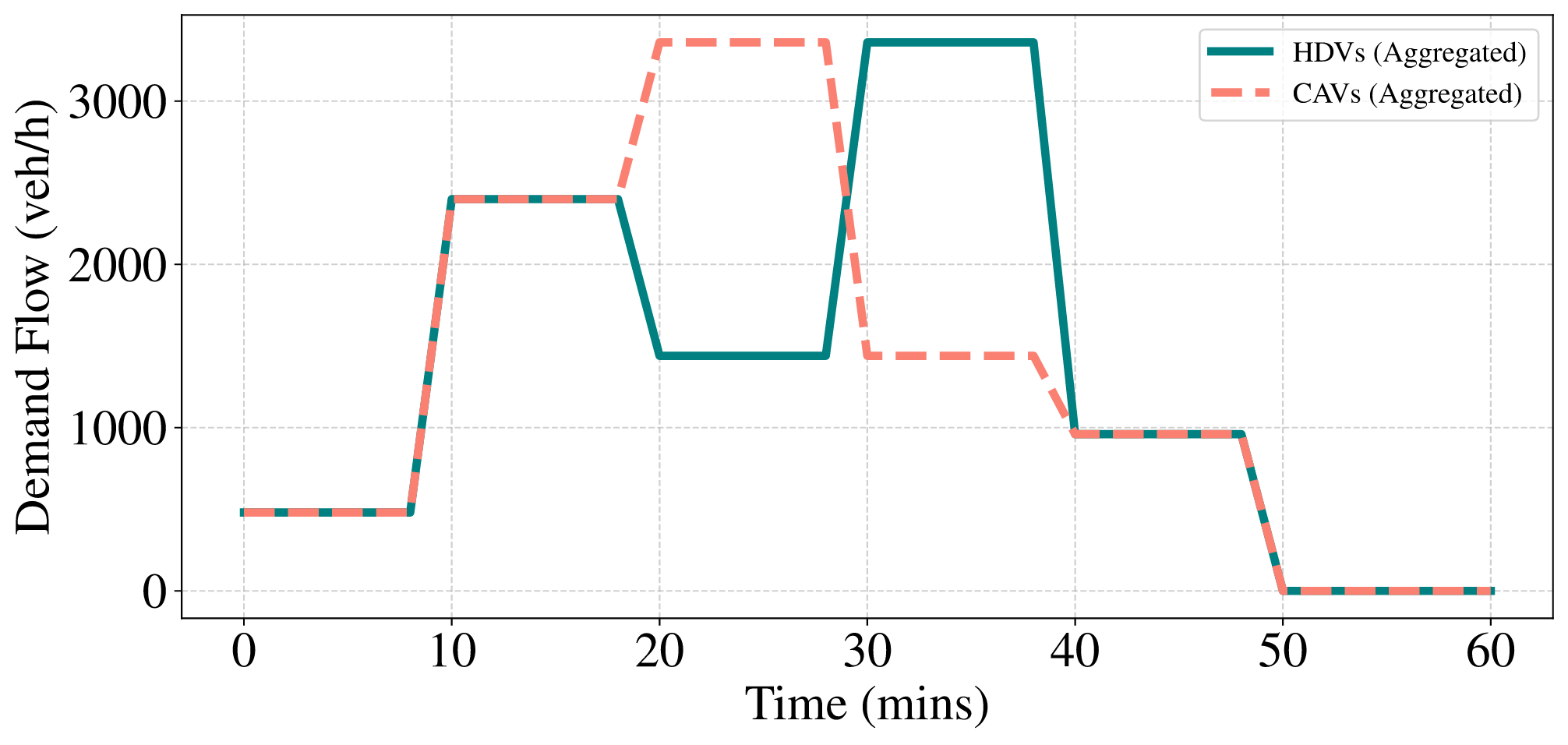}
  \caption{Aggregated time-varying demand profiles for CAVs and HDVs for testing robustness of DynamicSF strategy}
\label{fig:dem}
\end{figure}

\begin{figure*}[tb]
\centering\includegraphics[width=0.75\textwidth]{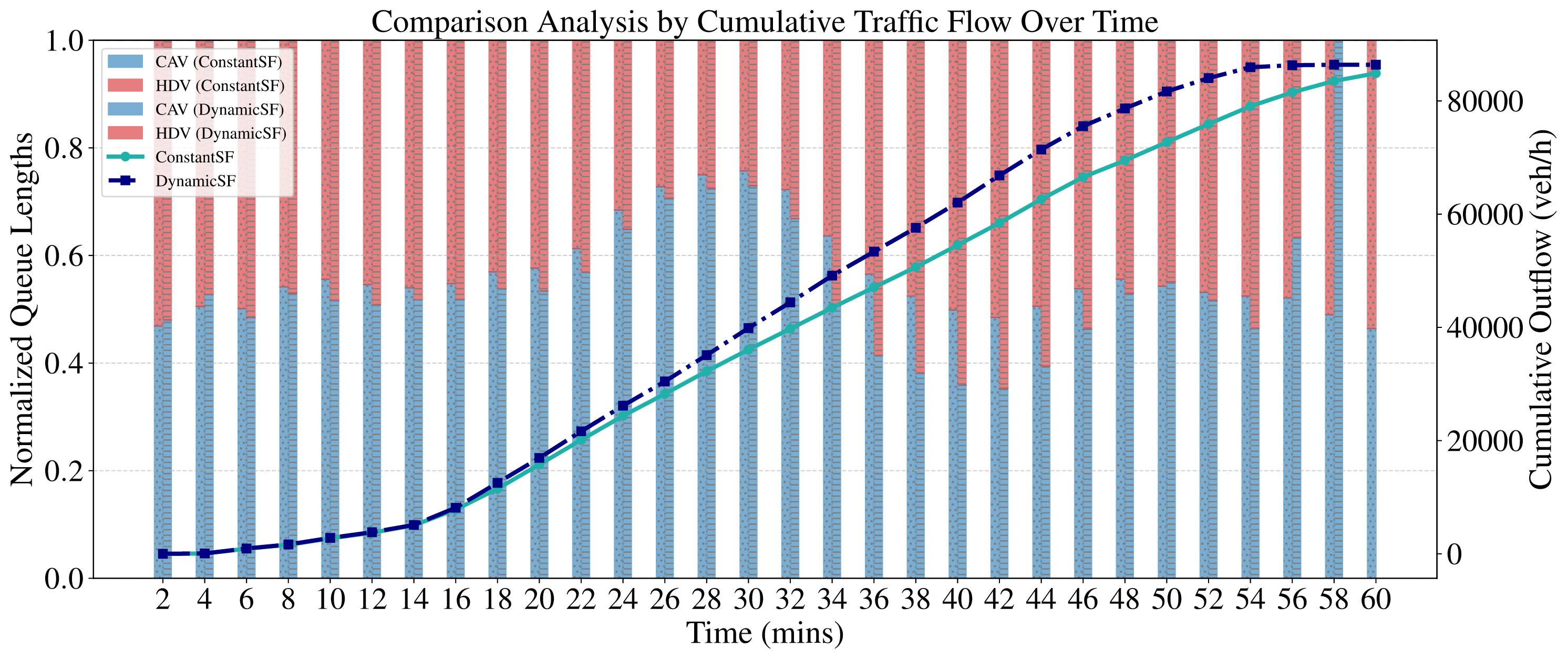}
\caption{Cumulative flow under ConstantSF and DynamicSF strategies across varying CAVs-HDVs proportions (illustrated by normalized queue lengths), evaluated with activation logic and a prediction horizon~\( K = 3 \).}
  \label{fig:rob_constdynam}
\end{figure*}

\subsection{Robustness Evaluation}
\label{robust_analysis}
This section tests the robustness of the DynamicSF strategy (implemented with activation logic) considering a demand with varying CAV--HDV proportions. The demand setup in Figure~\ref{fig:dem} is designed to reflect realistic mixed traffic conditions with distinct temporal flow patterns across vehicle classes, i.e. CAVs--HDVs. The objective is to evaluate the robustness and adaptability of the proposed control strategy, DynamicSF, under varying traffic intensities and compositions. In Figure~\ref{fig:dem}, the demand for CAVs is composed of origin-destination pairs that are connected with straight lines, i.e. vertically or horizontally with their respective pairing (e.g. $(1, 12), (8, 38)$), while for HDVs the demand is originated in all the entry links, thus resulting in proportional demand profile divided to all entry links for origins i.e. CAVs--HDVs: $\{50-50 , 70-30, 30-70, 50-50, 0-0\}$.

\begin{table}[tb]
  \centering
  \caption{Performance of the DynamicSF strategy considering varying CAVs and HDVs demand proportions}
  \label{tab:robust_analysis}
  \begin{tabular}{c *{6}{c}}
    \toprule
    \textbf{Prediction} 
    & \multicolumn{2}{c}{\textbf{TMQ (veh)}} 
    & \multicolumn{2}{c}{\textbf{ATT (h)}} 
    & \multicolumn{2}{c}{\textbf{Delay (s/km)}} \\
    \cmidrule(lr){2-3} \cmidrule(lr){4-5} \cmidrule(lr){6-7}
    \textbf{horizon}
    & \textbf{Avg} & \textbf{Std dev} 
    & \textbf{Avg} & \textbf{Std dev} 
    & \textbf{Avg} & \textbf{Std dev} \\
    \midrule
    2 & 109 & 7.6 & 255 & 7.0 & 116 & 7.7\\
    3 & 88 & 1.8 & 230 & 3.8 & 88 & 1.5 \\
    5 & 73 & 0.8 & 215 & 1.0 & 83 & 0.6 \\
    \bottomrule
    \vspace{0.0pt}
  \end{tabular}
\end{table}

The performance metrics for evaluating robustness are TMQ and ATT as introduced in Section~\ref{efftraffmet}. The results of the DynamicSF strategy are presented in Table \ref{tab:robust_analysis}, demonstrating that it adapts with varying CAVs presence in the network, resulting in a decrease of TMQ and ATT, proving overall efficiency and robustness. DynamicSF adapted seamlessly in various proportional divisions of CAVs--HDVs, maintaining consistent outflow, and thus providing the best results as per the evaluation of performance metrics. 

A comparative robustness analysis between the ConstantSF and DynamicSF strategies is presented in the form of cumulative flow, as shown in Figure~\ref{fig:rob_constdynam}. Both strategies are evaluated with activation logic framework with a prediction horizon of \( K = 3 \). The stacked bars represent normalized queue lengths (left y-axis) for CAVs and HDVs under both control strategies, while the line plots illustrate the cumulative traffic outflow (right y-axis, veh/h) over time. The normalized queue length illustrates the spatial and temporal imbalances in queue distribution. In the part with an equal proportion of vehicles (i.e., 50\% CAVs and 50\% HDVs), the performance of ConstantSF and DynamicSF is nearly identical, as the saturation flow rate remains effectively constant due to constant distribution of CAVs and HDVs. However, as the CAV penetration varies from 50\% to 30\%, and then to 70\%, DynamicSF exhibits a clear performance advantage over ConstantSF, highlighting its adaptability under dynamic traffic compositions. DynamicSF enhanced network throughput by enabling the use of shorter headways for CAVs, whereas ConstantSF uses fixed aggregate saturation rate. Notably, this performance gap persists even under conditions with nominally constant CAV penetration, particularly at higher demand levels. This is primarily due to the heterogeneous distribution of CAVs and HDVs throughout the network, especially in interlink segments, where vehicle proportions may fluctuate as a result of diverse turning movements and network topology.

\subsection{Sensitivity Analysis of Weight Parameters}
\label{weights_sensitivity}
\begin{figure*}[tb]
    \captionsetup{
        font=small, 
        labelfont=up, 
        textfont=up, 
        justification=centering}
    \captionsetup[subfloat]{
        font=small, 
        labelfont=up, 
        textfont=up,
        position=bottom}
    \centering
    \subfloat[\label{fig:weight1}]{\includegraphics[ width=0.39\textwidth,keepaspectratio]{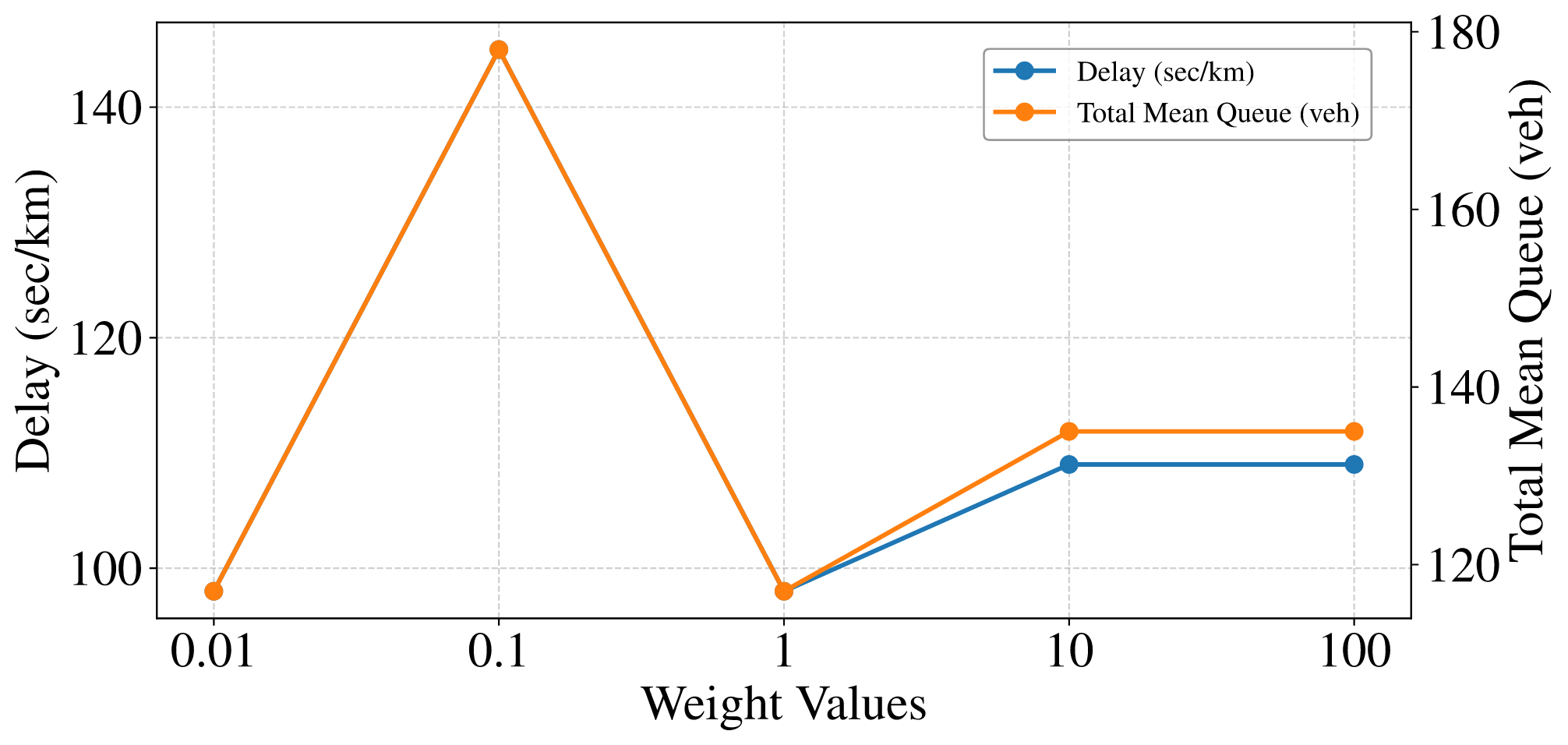}}\hspace{0.5cm}
    \subfloat[\label{fig:weight2}]{\includegraphics[ width=0.39\textwidth,keepaspectratio]{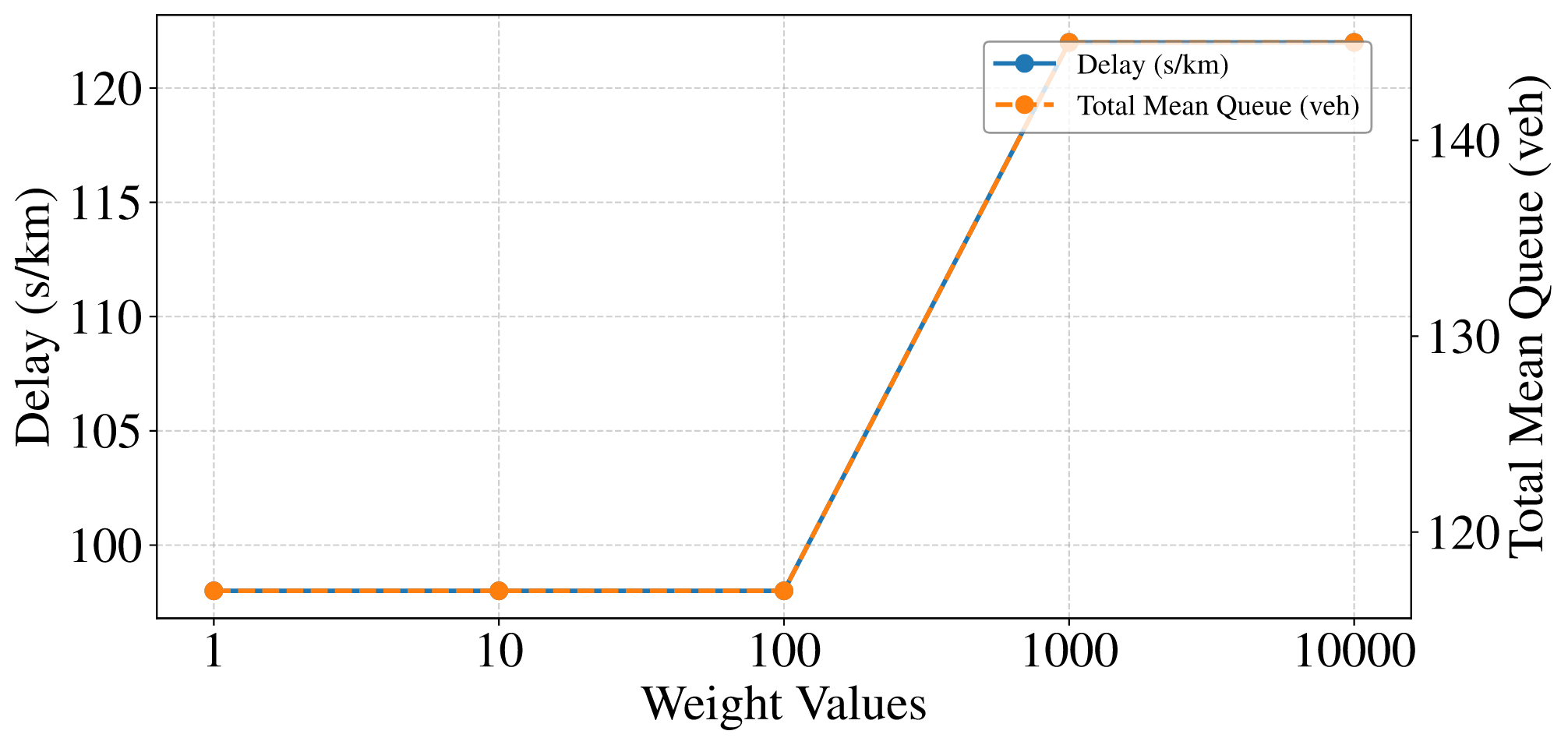}}
    \vspace{0.1cm}
    
    \subfloat[\label{fig:weight3}]{\includegraphics[ width=0.39\textwidth,keepaspectratio]{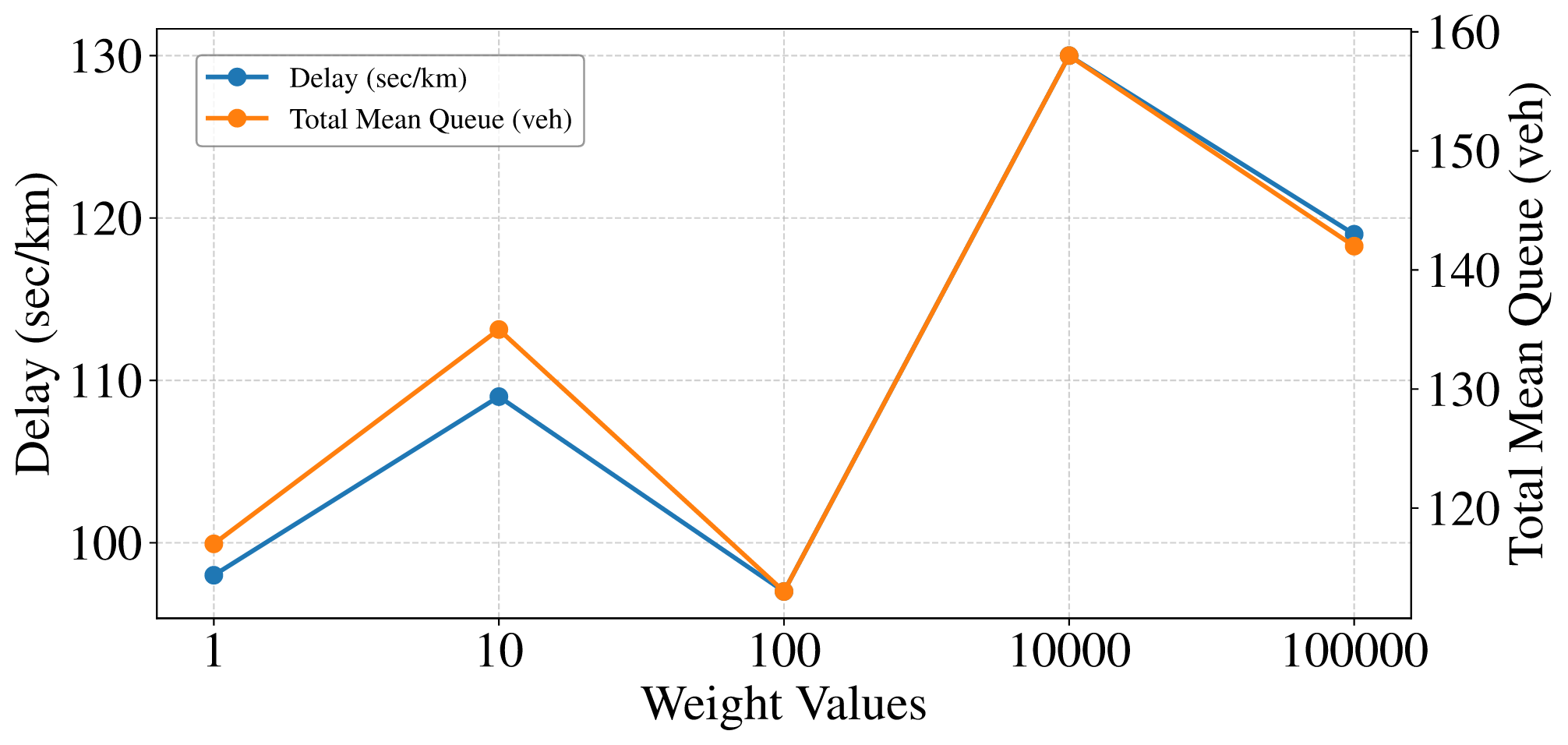}}\hspace{0.5cm}
    \subfloat[\label{fig:weight4}]{\includegraphics[ width=0.39\textwidth,keepaspectratio]{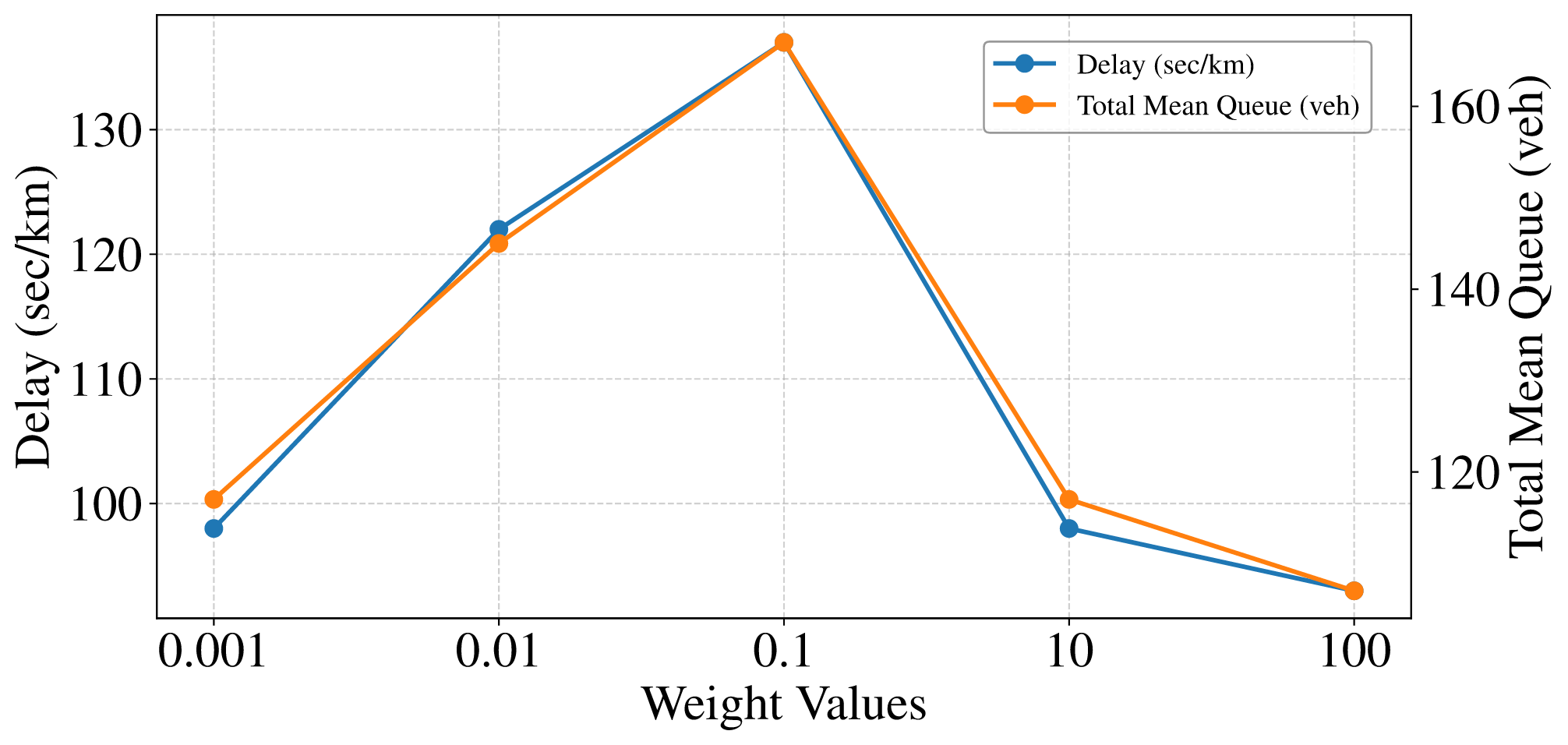}}
    \caption{Sensitivity analysis of the weights term of the objective function~\ref{eq:cost} (a) $\omega_1$, (b) $\omega_2$, (c) $\omega_3$, and (d) $\omega_4$.}
    \label{fig:weight_sense}
\end{figure*}

To evaluate the impact of the weighting parameters $\omega_1$, $\omega_2$, $\omega_3$, and $\omega_4$ in the multi-term objective function, a systematic sensitivity analysis is conducted. Each weight is varied individually over a range of values, while key performance indicators such as Delay (s/km) and Total Mean Queue, i.e. TMQ (veh), are monitored.  All simulations are performed using the demand profile from Table~\ref{tab:dem} and with $K = 3$, with each data point representing the average of $5$ replications to account for stochastic variability in traffic dynamics. The effects of these weight variations are illustrated in Figure~\ref{fig:weight_sense}.

The results exhibit distinct trends for each weight, highlighting optimal ranges where Delay and TMQ are minimized. For \(\omega_1\), which balances queue penalties between CAVs and HDVs, both metrics peak at 0.1 (Delay \(\approx\) 140 s/km, TMQ \(\approx\) 170 veh) before stabilizing at higher values for \(\omega_1 >\) 10 (Delay \(\approx\) 110 s/km, TMQ \(\approx\) 135 veh), suggesting that beyond a certain threshold the controller response becomes relatively insensitive to further increases in this weight. In contrast, \(\omega_2\), the terminal cost for CAVs routing, maintains better performance (Delay \(\approx\) 95 s/km, TMQ \(\approx\) 115 veh) from 1 to 100, but degrades for higher values, suggesting that excessive emphasis on terminal routing objectives can lead the controller to prioritize route completion at the expense of local traffic efficiency. For \(\omega_3\), penalizing larger queues, metrics reach a minimum around 100 (Delay \(\approx\) 95 s/km, TMQ \(\approx\) 110 veh) before increasing. The trend indicates that an overly strong penalty on queue magnitudes may distort the controller’s balance between short-term traffic progression and queue reduction. Finally, \(\omega_4\), the signal smoothing factor, shows peaks near 0.1 (Delay \(\approx\) 135 s/km, TMQ \(\approx\) 170 veh), with better performance at extremes but overall improvement for \(\omega_4 \geq \) 10. At smaller values of \(\omega_4\), the controller can change signal timings more freely, which may increase fluctuations in green times. At higher values of \(\omega_4\), excessive smoothing can limit the controller’s ability to react to changing traffic conditions, even though it reduces abrupt signal changes. Therefore, \(\omega_4\) should be selected to balance responsiveness and signal stability, as discussed in Section~\ref{sec:sign_perf}.

\subsection{Computational Performance}
\label{comp_time}
This subsection evaluates the average computational time per cycle (at each iteration of the  MPC controller) for the DynamicSF solution with prediction horizon $K = 2$. 
All computations are performed on a laptop equipped with an Intel Core i5-1135G7 processor (4 cores, 8 threads, 1.38 GHz base clock, up to 2.4 GHz boost), and 16 GB DDR4 RAM running Windows 10 64-bit.

\begin{table}[tb]
    \centering
    \caption{Average computation time per cycle for DynamicSF with~$K = 2$}
    \label{tab:comp_time}
    \begin{tabular}{c c c}
        \toprule
        \textbf{Number of Envelopes (N)} & \textbf{Avg (s)} & \textbf{Std dev (s)} \\
        \toprule
        5 & 1.78 & 1.15 \\
        7 & 2.68 & 1.65 \\
        9 & 4.91 & 4.21 \\
        \bottomrule
  \end{tabular}
\end{table}

Table~\ref{tab:comp_time} presents the average computation times and standard deviations for the DynamicSF mixed autonomy strategy evaluated at varying numbers of envelopes $N = 5$, $7$, and~$9$. The results show a near-linear increase in average computational time with increasing $N$, likely due to the growing number of variables and constraints that must be processed within each MPC cycle. The increasing standard deviation further suggests greater variability and reduced predictability of execution times at larger problem sizes.
These findings highlight a critical trade-off between computational efficiency and problem complexity. While the algorithm performs efficiently and consistently at lower $N$, the escalating average times and variability at higher $N$ suggest potential limitations for time-sensitive or large-scale applications. This degradation in performance stability underscores the need for optimization strategies, such as parallel processing or adaptive techniques \cite{schryen2020parallel}, to enhance robustness and ensure reliable deployment across a broader range of envelope counts.
Importantly, despite the observed increase in computation time and variability with higher $N$, the method remains well within the time bounds typically required for real-time traffic control applications. Therefore, the DynamicSF approach is suitable for real-time implementation.

\section{Conclusion}
\label{con}
This paper presented a novel traffic optimization framework tailored for mixed autonomy urban environments, integrating a dynamic saturation rate within an extended multi-commodity store-and-forward model to jointly manage CAVs and HDVs. By formulating the problem as a non-convex NQP and employing a piecewise MILP relaxation, the proposed DynamicSF strategy optimizes traffic signal timings and CAVs routing information (based on turning rates) in real-time through an MPC embedded approach. Microscopic simulations were conducted in Aimsun Next on a grid network of 16 intersections with 40 links, demonstrating significant performance improvements over the baseline ConstantSF (store-and-forward-based optimization not considering varying saturation rate) and FixedTime strategies. 
As with other traffic control models, the proposed framework represents an abstraction of real-world traffic dynamics. While it cannot fully capture the stochastic and behavioral characteristics of mixed traffic environments, it can represent key deterministic traffic-flow properties, such as network queue lengths, signal timing effects, and routing behavior of connected vehicles under controlled demand conditions. The proposed DynamicSF controller remains computationally efficient while accounting for varying demand levels of CAVs and HDVs, and it consistently outperforms the constant saturation flow rate model in simulations that incorporate stochastic traffic dynamics. Nevertheless, additional robustness evaluations are required before real-world deployment. These include testing under demand uncertainty, stochastic driver behavior models, and network disruptions such as incidents or temporary capacity reductions, as well as explicitly modeling interaction effects between manually driven vehicles and automated vehicles. 
The robustness and efficiency of the mixed autonomy DynamicSF strategy highlight its potential for practical deployment in future urban mobility systems. However, current assumptions, such as fixed cycle time and fixed vehicle headways, limit its adaptability to more complex conditions. Future research will focus on relaxing these constraints, testing the framework on heterogeneous intersection configurations, and integrating adaptive models to further enhance responsiveness. 
Additionally, exploring parallel processing techniques could mitigate the observed increase in computational variability at larger scales, ensuring scalability and stability for widespread implementation.
 
\appendices
\section{Floyd-Warshall Algorithm}
\label{app:floyd-warshall}
The Floyd-Warshall algorithm is a dynamic programming approach designed to compute the shortest paths between all pairs of arcs in a weighted graph. Unlike single-source shortest path algorithms such as Dijkstra's, which determine the shortest paths from one point to all others, Floyd-Warshall efficiently handles the all-pairs shortest path problem in a single execution. Therefore, It is convenient for dense graphs to accommodate arc weights, provided there are no negative cycles. The employed algorithm in this research work is listed as in Algorithm \ref{alg:floyd-warshall} to calculate the CAVs routing parameter~$F_{(z, d)}$. The algorithm iteratively improves an initial distance matrix, considering each arc as an intermediate point, to derive the shortest path distances at the end.
\begin{algorithm}[tb]
\caption{Floyd-Warshall Algorithm}
\label{alg:floyd-warshall}
\begin{algorithmic}[1]
\State \textbf{Input} Adjacency matrix $F$ of a weighted graph $ G = (Z, Z)$, and $F_{(z_{1}, z_{2})}$ represents the weight of each arc $(z_{1}, z_{2})$ whereas $z_{1} , z_{2} = \{1, 2, \ldots, Z\}$, or $\infty$ if no arc exists and $0$ for $z_{1} = z_{2}$.
\State \textbf{Output} A distance matrix $F$ where $F_{(z_{1}, z_{2})}$ is the shortest path distance from arc $z_{1}$ to arc $z_{2}$.
\State \textbf{Let} $n \gets |Z|$
\State \textbf{Initialize} $F^{(0)} \gets F$ {\footnotesize \Comment{Initial distance matrix as weighted matrix}}
\For{$k = 1$ to $n$}
    \For{$z_{1} = 1$ to $n$}
        \For{$z_{2} = 1$ to $n$}
 \State $F^{(k)}_{(z_{1}, z_{2})} \gets \min (F^{(k-1)}_{(z_{1}, z_{2})},F^{(k-1)}_{(z_{1}, k)} + F^{(k-1)}_{(k, z_{2})})$
        \EndFor
    \EndFor
\EndFor
\State \Return $F^{(n)}$ {\footnotesize \Comment{Final distance matrix with all-pairs shortest paths}}
\end{algorithmic}
\end{algorithm}

\section{Microscopic Simulation Parameters}
\label{app:sim_parameters}
Table~\ref{tab:sim_parameters_appendix} summarizes the microscopic simulation parameters used in Aimsun Next for our simulations. 
\begin{table}[tb] 
\centering 
\caption{Microscopic simulation parameters used in Aimsun Next} 
\label{tab:sim_parameters_appendix} 
\begin{tabular}{l c} 
\toprule 
\textbf{Parameter} & \textbf{Value} \\ 
\midrule HDV car-following model & Gipps \\ 
CAV car-following model & CACC \\ 
Vehicle length & 4 m \\ 
Vehicle width & 2 m \\ 
Network speed limit & 20 km/h \\ 
Maximum acceleration & 3 m/s$^2$ \\ 
Normal deceleration & 4 m/s$^2$ \\ Maximum deceleration & 6 m/s$^2$ \\ 
HDV route choice & Binomial stochastic routing \\ 
CAV route choice under MPC & Optimized by MPC \\ 
CAV route choice without MPC & Shortest-path routing \\ CACC speed gain & 0.0125 \\ 
CACC distance gain & 0.5 s$^{-1}$ \\ 
CACC leader time gap & 0.9 s \\ 
CACC follower time gap & 0.9 s \\
CACC platoon size & 0 \\
Simulation step & 0.1 s \\
CAV reaction time & 0.9 s \\
HDV reaction time & 1.8 s \\
\bottomrule 
\end{tabular} 
\end{table}

\section{Open-Loop Comparison Between NQP and MILP Approximations}
\label{app:openloop}
The optimization framework proposed in our work originates from the non-convex formulation~\eqref{eq:optProb}, which is characterized as a Non-Convex Quadratic Program (NQP). The non-convexity arises from the bilinear terms associated with the dynamic saturation rate~\eqref{eq:sat} and transport-flow vector~\ref{eq:transportflow}. As discussed in Section~\ref{sol}, these bilinearities render the problem computationally demanding. Therefore, we adopted a piecewise relaxation to reformulate the problem into a Mixed-Integer Linear Program (MILP).

We present here an analysis to assess the approximation quality as well as computational efficiency of the MILP relaxation of the problem against NQP formulation. The experiment considers the same network configuration as mentioned in section~\ref{exp} and demand profile from Table~\ref{tab:dem}, considering only an open-loop solution to the optimization problem. The evaluation horizon consists of $30$ cycles (with a cycle length of $120$~s), corresponding to a one-hour simulation period. Both the NQP and MILP formulations are solved using the Gurobi optimizer~\cite{gurobi} for consistency, with the latter evaluated using envelopes \(N \in \{5,7,9\}\).

To quantify the accuracy of the MILP relaxation relative to the original NQP formulation, the approximation error is computed using the relative deviation of the objective values:

\[
\text{Approx. error (\%)} =
\frac{\left|J_{\text{MILP}} - J_{\text{NQP}}\right|}
     {J_{\text{NQP}}} \times 100
\]

where $J_{\text{NQP}}$ denotes the objective function value obtained from the NQP formulation and $J_{\text{MILP}}$ denotes the objective function value obtained from the MILP approximation. The objective function cost is defined in~\eqref{eq:cost}. Table~\ref{tab:openloop} summarizes the computational time and objective values obtained for each formulation.

\begin{table}[tb]
\centering
\caption{Open-loop comparison between the NQP formulation and MILP approximations}
\label{tab:openloop}
\begin{tabular}{c c c c c}
\toprule
\textbf{Method} & \textbf{\makecell{Envelopes \\ (N)}} & \textbf{\makecell{Computation \\ time (s)}} & \textbf{\makecell{Objective \\ value}} & \textbf{\makecell{Approx. \\ error (\%)}} \\
\toprule
NQP & -- & 3835.76 & 835.85 & -- \\
\midrule
MILP & 5 & 97.56 & 821.27 & 1.75 \\
     & 7 & 298.46 & 825.26 & 1.27 \\
     & 9 & 408.35 & 836.24 & 0.05 \\
\bottomrule
\end{tabular}
\end{table}

While McCormick envelopes alone provide a lower bound for a minimization problem, the MILP here also includes a piecewise discretization as described in~\ref{eq:sat_convex}, which slightly restricts the feasible region. Therefore, the objective function value of MILP approximation can exceed the NQP solution, as seen for $N=9$.

It is clear to note that solving the original NQP formulation is significantly more demanding in terms of computational time, contrasting to the MILP approximations which can be solved considerably faster even with higher envelope segments. Moreover, the approximation quality of MILP method improves as the number of envelopes increases. With $N=9$, the MILP solution closely matches the NQP solution, achieving an approximation error of only $0.05\%$. 


\bibliographystyle{IEEEtran}
\bibliography{ieeeabrv, bibliography}
\newpage

\begin{IEEEbiography}[{\includegraphics[width=1in,height=1.25in,clip,keepaspectratio]{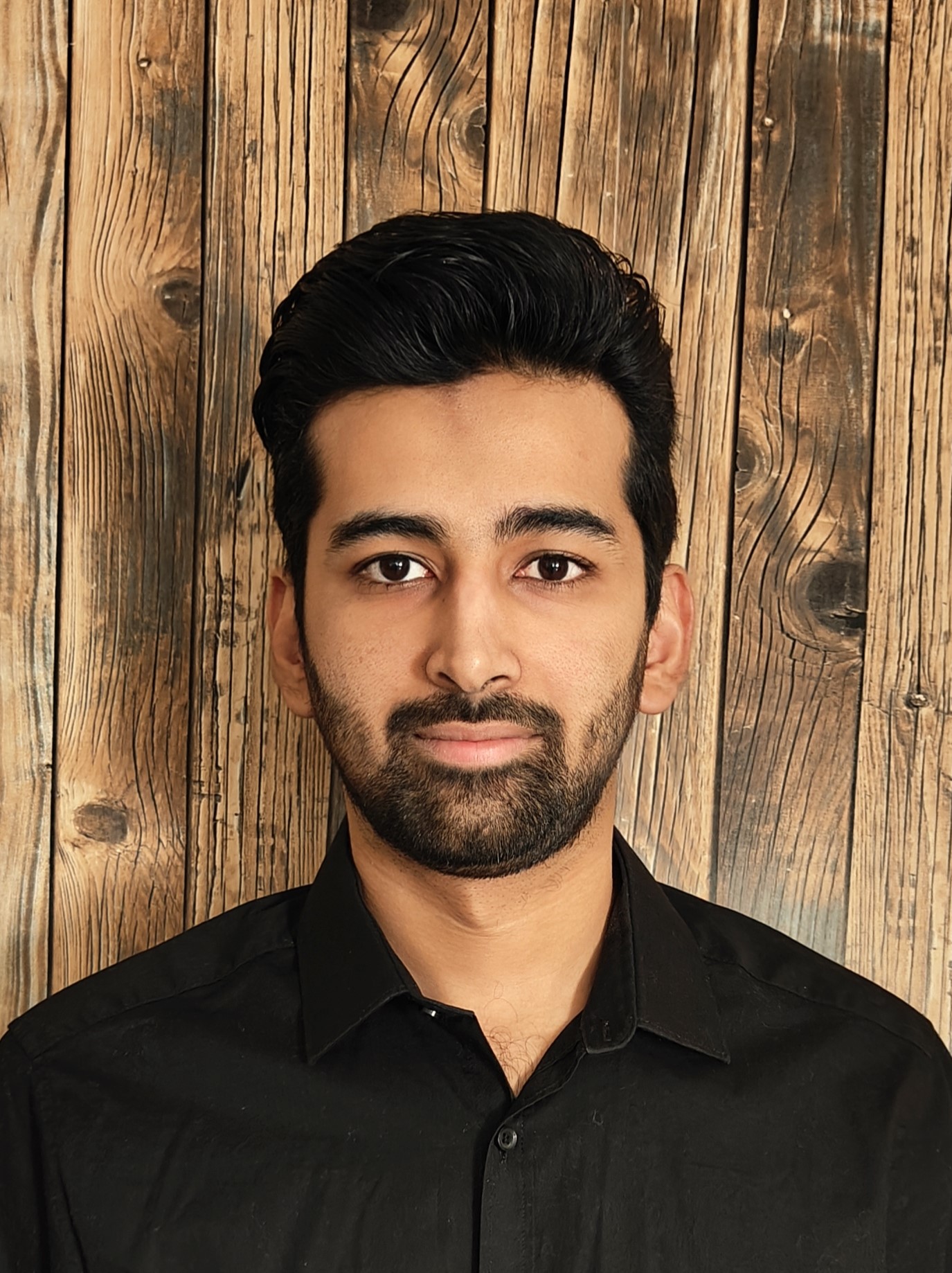}}]{Haris Muhammad } is a Doctoral Researcher at the Department of Built Environment, Aalto University, Finland, where he is pursuing a Doctor of Science (Technology) in the School of Engineering. He previously obtained his Master’s degree in Electrical and Computer Engineering from King Abdullah University of Science \& Technology, Saudi Arabia, and his Bachelor’s degree in the same field from the National University of Sciences \& Technology, Pakistan.
Prior to his doctoral studies, he worked as a Graduate Research Assistant at Robotics, Intelligent Systems, and Control Lab, King Abdullah University of Science \& Technology.
His current work focuses on developing model-based solutions for complex urban mobility challenges. His broader interests include real-time traffic management systems, air traffic management, applied optimization, mathematical modeling, and intelligent transport systems.
\end{IEEEbiography}

\vspace{11pt}

\begin{IEEEbiography}[{\includegraphics[width=1in,height=1.25in,clip,keepaspectratio]{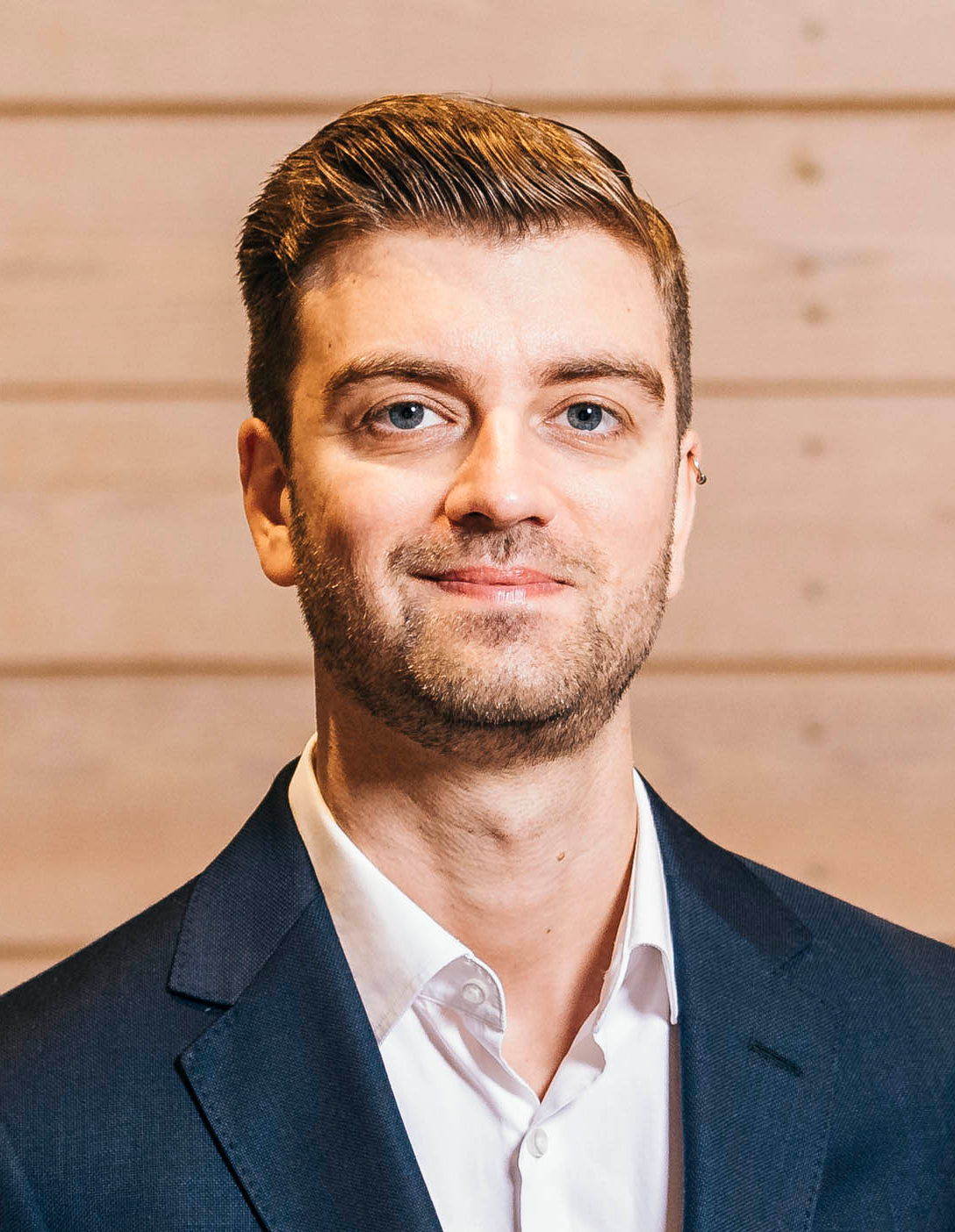}}]{Claudio Roncoli } is an Associate Professor at the Division of Mobility and  Industrial Management, KU Leuven, Belgium. Before joining KU Leuven, he was an Associate Professor at Aalto University, Finland, where he still holds a Visiting Professor position. Previously, he was a researcher at the Technical University of Crete, Greece, and the University of Genova, Italy, as well as a visiting researcher at the University of Queensland and the University of Sydney, Australia, and at the Imperial College London, UK. 
Claudio has been involved in several national and international research projects as a principal investigator. His research interests include real-time traffic management; modeling, optimisation, and control of traffic systems with connected and automated vehicles; as well as smart mobility and intelligent transport systems.
\end{IEEEbiography}

\vfill

\end{document}